\documentclass[aps,floatfix,groupedaddress,longbibliography,amsmath,amssymb,pre,preprint]{revtex4-1}
\usepackage[utf8]{inputenc}
\usepackage{graphicx}

\newcommand{\erf}{\operatorname{erf}}

\begin{document}

\title{Fourth-order analysis of a diffusive lattice Boltzmann method for barrier coatings}
\author{Kyle T.~Strand}
\email{\texttt{kyle.t.strand@ndsu.edu} (corresponding author)}
\author{Aaron J.~Feickert}
\email{\texttt{aaron.feickert@ndsu.edu}}
\author{Alexander J.~Wagner}
\email{\texttt{alexander.wagner@ndsu.edu}}
\affiliation{Department of Physics, North Dakota State University, NDSU Dept 2755, PO Box 6050, Fargo ND 58108-6050, USA}

\date{\today}

\begin{abstract}
We examine the applicability of diffusive lattice Boltzmann methods to simulate the fluid transport through barrier coatings, finding excellent agreement between simulations and analytical predictions for standard parameter choices. To examine more interesting non-Fickian behavior and multiple layers of different coatings, it becomes necessary to explore a wider range of parameters. However, such a range of parameters exposes deficiencies in such an implementation. To investigate these discrepancies, we examine the form of higher-order terms in the hydrodynamic limit of our lattice Boltzmann method. We identify these corrections to fourth order and validate these predictions with high accuracy. However, it is observed that the validated correction terms do not fully explain the bulk of observed error. This error was instead caused by the standard finite boundary conditions for the contact of the coating with the imposed environment. We identify a self-consistent form of these boundary conditions for which these errors are dramatically reduced. The instantaneous switching used as a boundary condition for the barrier problem proves demanding enough that any higher-order corrections meaningfully contribute for a small range of parameters. There is a large parameter space where the agreement between simulations and analytical predictions even in the second-order form are below 0.1\%, making further improvements to the algorithm unnecessary for such an application.
\end{abstract}

\maketitle

\section{Introduction} \label{sec:intro}
Coating systems are used heavily in industry for the protection of materials and infrastructure. Common examples include the paints on cars, bicycles, and houses; the layered coating systems used on boats and airframes; and the coatings used to protect bridges. In all cases, the goal of the coating system is to protect the underlying substrate from ingress by aggressive particulate, gaseous, or fluid materials while remaining aesthetically intact.

Crosslinked polymer networks, also called thermosets, are typically chosen in protective applications due to their net-like structure. In such a structure, precursor materials are chemically bonded through a crosslinking and curing process to form a three-dimensional structural network. This network acts as a physical and chemical barrier that attempts to prevent permeation by water, salt, particulate matter, and other environmental contaminants. Since substrates are often materials susceptible to corrosion, like aluminum or steel, it is essential that moisture not be permitted to reach the substrate in appreciable quantities.

As is known, most coatings permit, to some degree, moisture ingress \cite{Taylor2012169}. This can be due to imperfections in the preparation process \cite{Kroll201582}, the formation of void space during curing or cooling \cite{Zee201555}, or because of damage in service. To help detect coating formulations that may be unsuitable for use in the field, accelerated weathering testing is used to determine failure rates and modes in the lab. Much research has been devoted to the relationship between accelerated testing, comparable real-world testing, and service life, but no complete and predictive model exists that accurately correlates a coating's performance in lab testing, performance in field testing, and failure modes or lifespan that is likely to occur in service \cite{doi:10.1021/bk-2002-0805.ch001}.

Since moisture entering the coating is conserved, the dynamics of fluid density $\rho$ obey the continuity equation \begin{equation}
\partial_t \rho + \nabla\mathbf{j} = 0,
\end{equation}
where $\mathbf{j}$ is the mass current. Assuming an isotropic coating, mass current will be in the direction of negative density current. We denote the proportionality between the current and negative gradient by $D$, which in the simplest case is a constant. Later, we consider a more general $D(\rho)$. We therefore have $\mathbf{j} = -D\nabla\rho$. With this constitutive relation for the mass current, we recover the well-known diffusion equation.

Several methods exist to model idealized diffusion. Early work focused primarily on precise mathematical modeling and numerical solutions to boundary-matched differential equations governing diffusion \cite{crank1979mathematics,de1998monitoring}. Modern approaches include network connectivity models \cite{AIC:AIC690480104}, Monte Carlo simulations \cite{SAHIMI19912225}, and finite-element analysis \cite{Mu2007} for more complex structures like porous media where an effective diffusivity is desired. However, approaches dealing with pore structures may depend on the structure and porosity of the material in question, quantities that most often unknown \textit{a priori}. Additionally, finite-element models tend to be computationally complex and often rely on commercial closed codes. As a whole, there is comparatively little known about the precise dynamics of diffusion through polymeric coatings \cite{Pathania2017149}.

Additionally, different approaches exist when multiple layers are considered. In the case of multiple hydrophobic barrier coatings, boundary-matching Fickian solutions can be used \cite{de1998monitoring} and matched to experiment via electrochemical methods. When a base coating is hydrophilic, as is the case with some primers, an alternative approach couples Fickian diffusion for any overlying barrier coatings with the assumption of an instantaneous reservoir for the base layer \cite{Baukh20123304}. In either case, different coatings in a multi-layer stackup differ in their effective diffusivity.

Any numerical technique used to model the progression of moisture in such a stackup must stably account for a wide range of diffusion constants. Since laboratory testing of candidate barrier coating systems typically includes cyclic exposure to moisture and dry ambient air over long periods of time, simulations of cyclic processes must maintain numerical stability over correspondingly longer time scales.

In this paper, we use lattice Boltzmann numerical techniques to determine the accuracy of modeling moisture ingress through a finite coating system exposed to a reservoir and adhered to an ideal substrate. Because of the necessity of modeling a wide range of saturation levels and diffusivity in the case of multi-layer systems, we analyze the error introduced in the traditional second-order approximation to the diffusion equation used in lattice Boltzmann approaches. To investigate the nature of this error, we introduce a fourth-order correction and perform a Fourier component analysis to confirm the correctness of our results. We show that the bulk of the second-order error in such a system arises from the boundary conditions used, and comment on the proper use of periodic systems to remove this error. Applications to multi-layer systems with variable diffusivity are discussed in the context of our analysis.

\section{Lattice Boltzmann methods} \label{sec:lb}
The lattice Boltzmann approach models densities $f_i$ defined on a discrete lattice space associated with discrete lattice velocities $v_i$. After being displaced to a new lattice position $x+v_i$, the densities at each lattice point are redistributed in a collision step. This method has been used extensively to model hydrodynamic behavior \cite{qian1992lattice,PhysRevLett.75.830,PhysRevLett.56.1505}, diffusion \cite{wolfdiffusion, shan}, electrostatics \cite{electrostatics}, and similar systems with high accuracy and computational efficiency. Notably, the hydrodynamic partial differential equations underlying such systems are not the starting point for the method, but rather emerge from it. Choices like the number of quantities conserved in the collision allow for the freedom of recovering the governing equations for a variety of different systems, as mentioned above.

A popular collision term defines a local equilibrium $f_i^0$ that only depends on the conserved quantities and then relaxes the actual density towards the local equilibrium. In this form the lattice Boltzmann equation can be written as
\begin{equation}
f_i(x+v_i,t+1) = f_i(x,t) + \sum_j \Lambda_{ij} \left[ f_j^0(\rho(x,t)) - f_j(x,t) \right].
\label{eqn:lb}
\end{equation}
Here $f_j^0$ is the local equilibrium density, $\Lambda_{ij}$ is a collision matrix, and $\rho(x,t)$ is the local density of the system, given by
\begin{equation}
\rho(x,t) = \sum_i f_i(x,t).
\end{equation}
The form of the collision matrix allows for further control of the algorithm, but this freedom is not explored in this paper. Most examples where this freedom has shown to be useful relate to simulations of hydrodynamic systems with very low viscosity. Such low viscosities may give rise to instabilities that can be controlled by a careful choice of the collision matrix. For diffusive systems like the one considered here, the advantages of utilizing multiple relaxation times are less well established (See Ginzburg \cite{PREGinzburg}), so we will employ the particularly simple collision matrix \begin{equation}
\Lambda_{ij} = \frac{1}{\tau}\delta_{ij}
\end{equation}
that was originally proposed by Qian \cite{qian1992lattice}, using a single relaxation time $\tau$.

It is necessary to impose moments on the equilibrium distribution, following the method of \cite{wolfdiffusion}. While not considered here, formulations of this method in the case of multiple components \cite{shan} and multiple relaxation times \cite{LiPRE2014,LePRE2015}. In particular, we impose the following (non-unique) moments on the distribution:
\begin{eqnarray}
\sum_i f_i^0 &=& \rho \label{eqn:momentrho}\\
\sum_i f_i^0v_{i\alpha} &=& 0 \label{eqn:momentmom} \\
\sum_i f_i^0v_{i\alpha}v_{i\beta} &=&  \rho\theta\delta_{\alpha\beta} \label{eqn:momentstress}
\end{eqnarray}
where the Greek indices are spatial dimensions and follow the Einstein notation.

Local density conservation is assured by Eqn. (\ref{eqn:momentrho}), while Eqn. (\ref{eqn:momentstress}) introduces a spatially uniform imposed temperature $\theta$. Following \cite{PhysRevE.94.033302}, a second-order Taylor approximation using this choice of moments leads to the lattice diffusion equation
\begin{equation}
\partial_t\rho = \nabla_\alpha\left(\tau - \frac{1}{2}\right) \nabla_\alpha(\rho\theta) \label{eqn:latticediff}
\end{equation}
and, if the temperature is constant, this recovers a diffusion equation with the diffusion constant
\begin{equation}
D = \left(\tau - \frac{1}{2}\right)\theta. \label{eqn:D}
\end{equation}

Testing for coating applications usually applies moisture somewhat homogeneously on the sample, either in soak testing or weathering chambers; drying also proceeds evenly. This reduces the problem of interest to an effectively one-dimensional case. For simulations, we use the simplest one-dimensional lattice Boltzmann model with the velocities $\{v_i\} = \left\{ 0, 1, -1 \right\}$. This one-dimensional lattice with the given velocities is known as a D1Q3 scheme. For this implementation of a diffusive system, the local equilibrium distribution can be written as
\begin{equation}
f_i^0 = \rho w_i,
\label{eqn:eqdist}
\end{equation}
where $w_i$ are the weights related to the magnitude of the velocities $\{v_i\}$. To recover the necessary moments, the weights are
\begin{eqnarray}
    w_0 &=& 1 - \theta \nonumber\\
    w_1 &=& \frac{\theta}{2} \nonumber\\
    w_2 &=& \frac{\theta}{2}.
\end{eqnarray}
The D1Q3 implementation then allows for a full and self-contained simulation method for a diffusive system.

\section{Application to water content of coatings} \label{sec:coating}
We wish to model the wetting of a single-layer coating via Fickian diffusion. Since coatings are frequently examined in the laboratory on test panels using weathering chambers that subject the coating to moisture, we will consider the case where the coating, represented by a lattice from $0 \leq x \leq L_x$, is exposed to a reservoir of varying concentration $\rho^b(t)$ at $x = 0$ and an impermeable substrate at the right end of the simulation lattice. The meaning of $\rho^b$ is the amount of water that will be absorbed just inside the coating as it is exposed to the environment. For an immersion in water, this corresponds to the maximal water content the coating can absorb, and we scale the density so that this value corresponds to $\rho=1$.  

\begin{figure}
    \centering
    \includegraphics[width=0.5\columnwidth,clip=true]{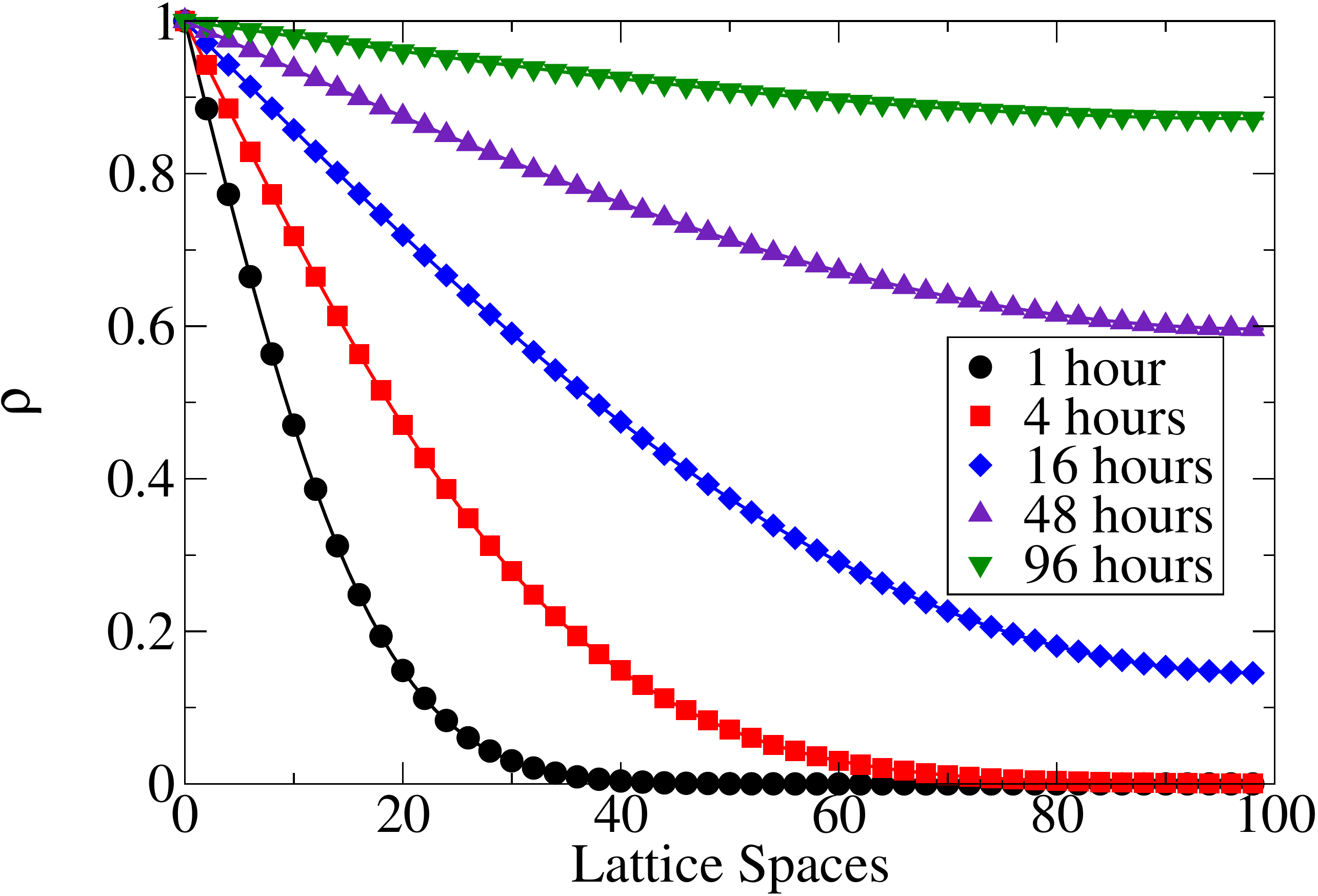}
    \caption{Concentration profile at $\tau = 1$, $\theta = 0.5$, $\rho_0=1$ at various times (symbol), with analytical solution $\rho(x,t)$ (solid line).}
    \label{fig:exposure}
\end{figure}

We must account for these two boundary conditions in our numerical simulation. We implement the source term by setting \begin{equation}
f_i(0,t) = f_i^0(\rho^b(t))
\end{equation}
and by replacing the streaming step at the right end by a bounceback algorithm, where the right-moving $f_1(L_x)$ is reinserted as an $f_2$ in the streaming step. The result for a step function $\rho^b(t) = \Theta(t)$ in the exposure is shown in Fig. \ref{fig:exposure}. We used a system with $L_x = 100$ lattice points, $\tau = 1$, $\theta = 0.5$, $\rho_0 = 1$, and ran the simulation for a variable number of iterations $T$. As expected, moisture is at first located closely to the surface and then penetrates the sample.

To verify the correctness of the simulation results, we construct an analytical solution for the concentration over time, using linear combinations of the well-known error function solution \cite{crank1979mathematics}. These are solutions of the diffusion equation for the initial condition of a step function in an infinite system. If the initial step goes from $2\rho_0$ to zero, then the solution is
\begin{equation}
\rho^{th,1}(x,t) = \rho_0\left( 1 - \erf\left(\frac{x}{\sqrt{4Dt}}\right)\right).
\end{equation}
This solution has a fixed point $\rho_{\operatorname{bath}}(0,t) = \rho_0$ at $x=0$, which corresponds to our boundary condition. So $\rho^{th,1}(x,t)$ for $x\geq 0$ and $t>0$ is the analytical solution for an infinite dry coating exposed to a reservoir starting at time $t=0$. Note that the long-time behavior gives $\rho^{th,1}(x,t \to \infty) = \rho_0$ as expected.

Suppose now that we have a finite one-dimensional coating extending from $0 \leq x \leq L_x$. At $x = 0$, the coating is exposed to a reservoir with fixed concentration $\rho(x=0,t) = \rho_0$. At $x = L_x$ is an impermeable substrate where $\nabla\rho(x=L_x,t) = 0$.

To account for the vanishing current at the substrate, we use an image source reservoir at $x = 2L_x$. This will ensure a vanishing gradient at $x = L_x$ and, by symmetry, a vanishing current. However, when the reflected concentration becomes nonzero at the reservoir again, we must subtract another image source reservoir at $x = -2L_x$ to maintain the correct boundary condition. Repeating this process infinitely, we arrive at the final solution that includes both reservoir and substrate:
\begin{equation}
\rho^{th}(x,t) = \rho_0\sum_{i=0}^\infty (-1)^i\biggl[ 2 + \erf\left(\frac{x-2(i+1)L_x}{\sqrt{4Dt}}\right) - \\ \erf\left(\frac{x+2iL_x}{\sqrt{4Dt}}\right) \biggr] \label{eqn:rho}
\end{equation}
For practical purposes, we find ten terms of the infinite sum in Eqn. (\ref{eqn:rho}) are entirely sufficient for most cases.

It is instructive to determine the correspondence between these numerical parameters and a laboratory case. A typical barrier coating might have thickness $X = 50\ \mu$m, diffusion constant in water $D \sim 10^{-14}\ \text{m}^2/\text{s}$, and be exposed to moisture in a weathering chamber for $T = 4$ hours at a time for testing. We can introduce reduced time, length, and density scales $t',x',\rho'$ such that
\begin{eqnarray}
t &=& Tt' \\
x &=& Xx' \\
\rho &=& \rho_0\rho'
\end{eqnarray}
and $0 \leq \{t',x',\rho'\} \leq 1$. Since the unit relationship $T = X^2/D$ holds by dimensional analysis, for any given experimental setup the quantity 
\begin{equation}
F \equiv \frac{TD}{X^2}
\end{equation}
is dimensionless and we have the scaled diffusion equation $\partial_{t'}\rho' = -\nabla_{x'}F\nabla\rho'$. Using the experimental parameters suggested above gives $F = 5.76 \times 10^{-2}$. In our simulations, we use total length $X = L_x = 100$ lattice sites, reservoir concentration $\rho_0 = 1$, $\theta = 0.5$, and $\tau = 1$. Since this gives a time scale $T \approx 2300$ iterations, this means one hour of equivalent macroscopic exposure corresponds to approximately $575$ simulation iterations. Further, the choice of $\tau = 1$ yields immediate relaxation of local distributions, so we would expect excellent agreement to theory.

\begin{figure}
    \centering
    \includegraphics[width=0.5\columnwidth,clip=true]{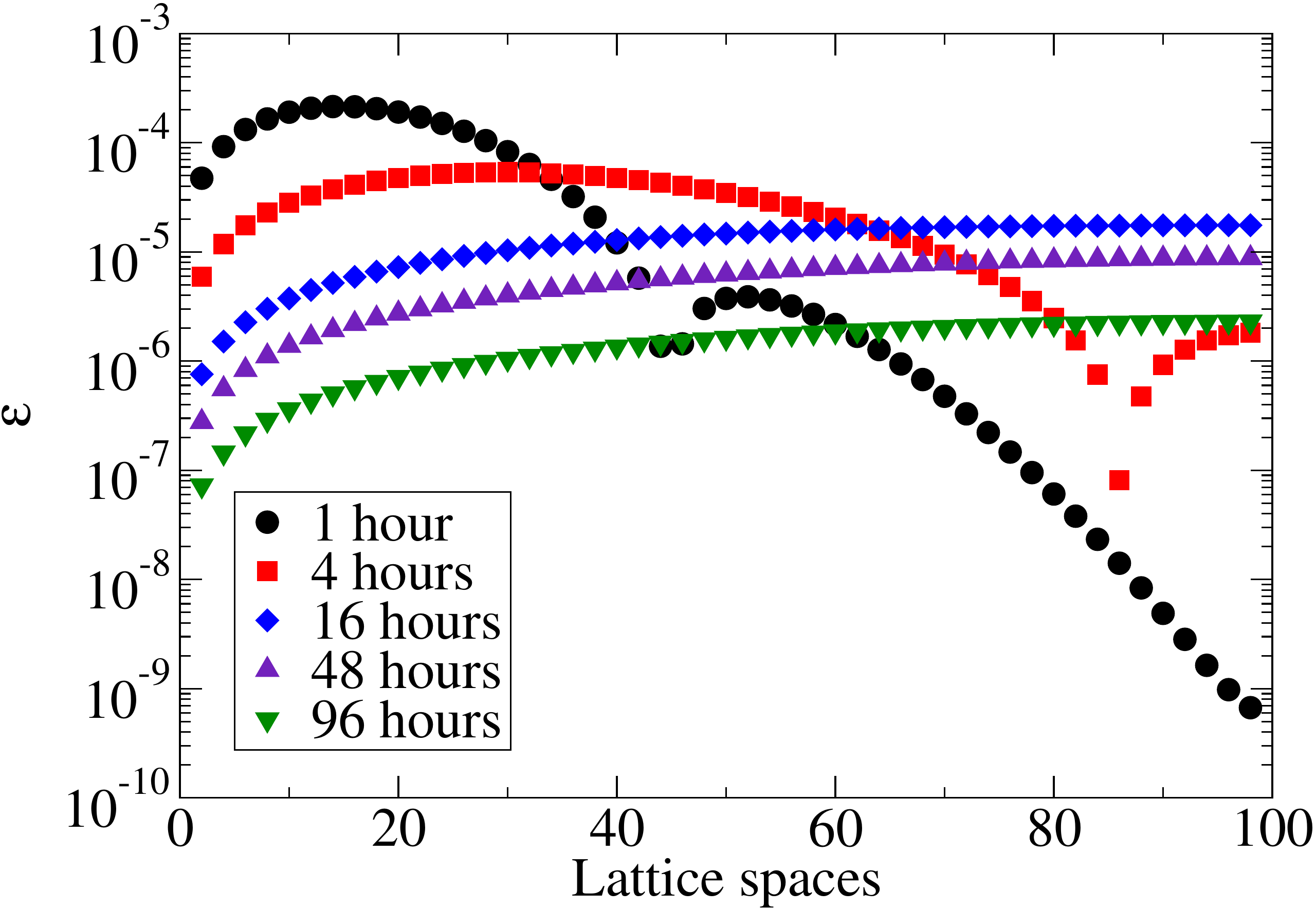}
    \caption{Absolute error profile $\epsilon$ between numerical and analytical concentration for exposure over time.}
    \label{fig:error-exposure}
\end{figure}

We are now in a position to comment on the accuracy of this simulation method in comparison to the analytical solution $\rho^{th}$ in Eqn. (\ref{eqn:rho}). For each of the exposure times in Fig. \ref{fig:exposure}, we compute the absolute error
\begin{equation}
\epsilon(x) \equiv \left|\rho(x,t) - \rho^{th}(x,t)\right|
\end{equation}
across the lattice space profile. The result is plotted logarithmically in Fig. \ref{fig:error-exposure}, showing excellent agreement. It is interesting to observe how the error changes over time; initially, the error drops substantially since moisture has not yet permeated through the entire coating lattice. This tail increases as the entire lattice becomes wet, but then uniformly decays as the numerical solution approaches saturation and agrees with the corresponding analytical solution.

While this method provides efficient and stable numerical modeling of a single coating, a given coating system might consist of two or more barrier layers in a stackup, each with a different diffusion constant that permits moisture ingress and egress at different rates from its neighbors. To extend this method to the simplest multi-layer case, we might wish to model a two-layer stackup consisting of idealized barrier coatings with different physical properties. To do so, our reservoir model is modified slightly, with the outer barrier coating represented at lattice sites $0 \leq x \leq L_x/2$ and the inner barrier coating at $L_x/2 \leq x \leq L_x$. Since the diffusion constant is controlled by the parameter $\tau$ in Eqn. (\ref{eqn:latticediff}), the presence of two diffusion constants requires that $\tau$ be position-dependent:
\begin{equation}
\tau = \tau(x) \equiv \left\{
\begin{array}{llr}
\tau_{\operatorname{out}} & , & 0 \leq x \leq L_x/2 \\
\tau_{\operatorname{in}}  & , & L_x/2 \leq x \leq L_x
\end{array}\right.
\end{equation}
Incidentally, changing the value of $\theta$ between the two regions will lead to different maximum water uptake in the layers, an important relationship that will be explored elsewhere.

\begin{figure}
    \centering
    \includegraphics[width=0.5\columnwidth,clip=true]{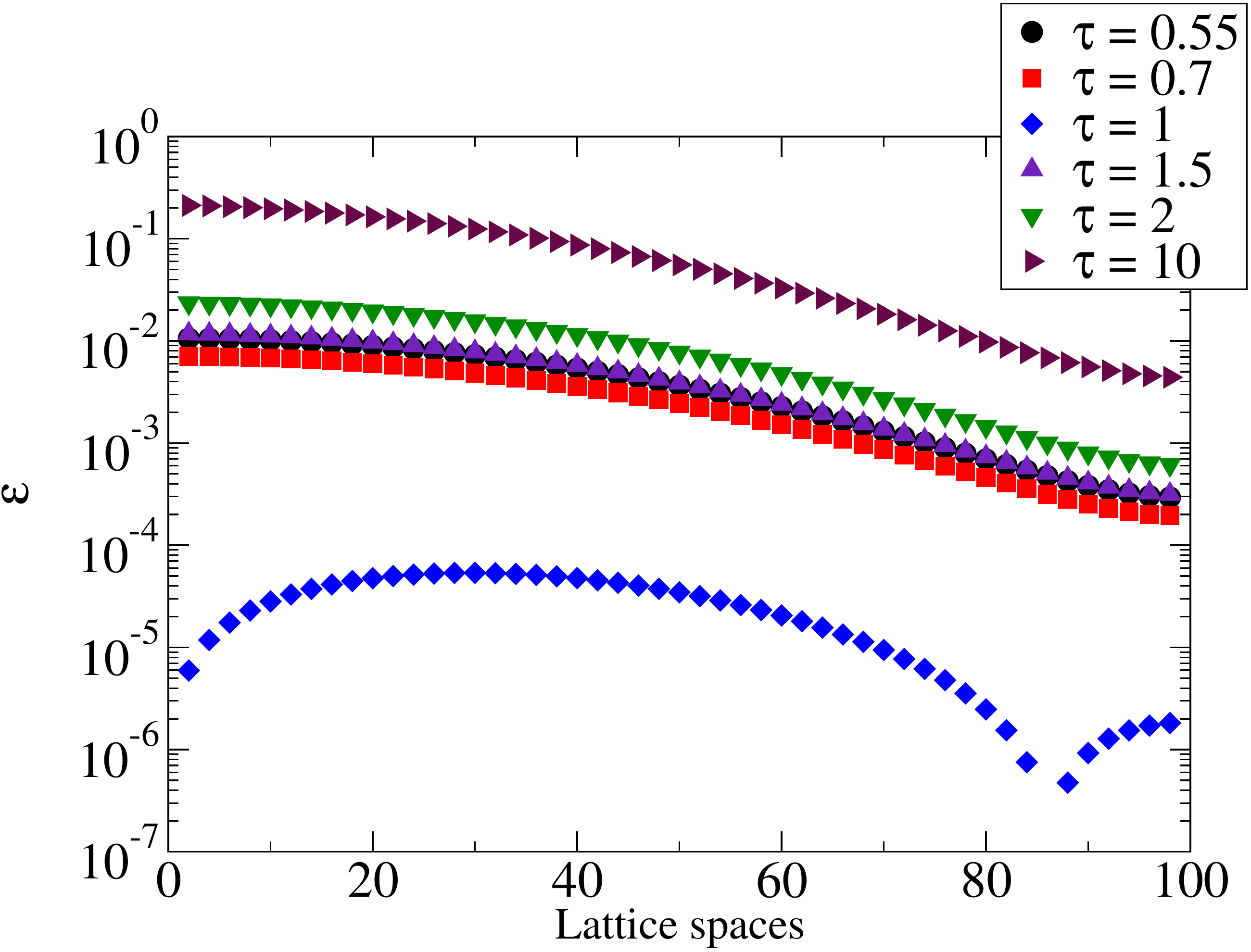}
    \caption{Absolute error profile $\epsilon$ between numerical and analytical concentration at various $\tau$. All simulations were run to the same scaled time, corresponding to four hours of macroscopic equivalent time.}
    \label{fig:error-finite}
\end{figure}

Although such a two-layer system is not investigated in this paper, it is essential to determine the range of $\tau$ values for which numerical and analytical solutions agree sufficiently over time. For efficient simulations, it is advantageous to choose $\tau$ as large as feasible, since this corresponds to a large diffusion constant and hence a shorter simulation time. For a quick initial evaluation, we run a series of lattice Boltzmann simulations with varying values of $\tau$ to the same macroscopic equivalent time of four hours of exposure. After that time, we compute the absolute error $\epsilon$ between numerical and analytical solutions across the entire lattice profile. Results are shown in Fig. \ref{fig:error-finite}. The choices of $\tau$, along with the corresponding time scale $T$, are shown in Table \ref{table:D}.

\addtolength{\tabcolsep}{6pt}
\begin{table}
\centering
\begin{tabular}{lllr}
$\tau$ & $\theta$ & $D$ & $T$ \\
\hline
0.55 & 0.5 & 0.025 & 23040 \\
0.70 & 0.5 & 0.10 & 5760 \\
1.0 & 0.5 & 0.25 & 2304 \\
1.5 & 0.5 & 0.50 & 1152 \\
2.0 & 0.5 & 0.75 & 768 \\
10.0 & 0.5 & 4.75 & 121
\end{tabular}
\caption{Values of $\tau$ and $\theta$ used in simulations, with corresponding diffusion constant $D$ and time scale $T$ corresponding to four hours of macroscopic equivalent exposure with $F = 5.76 \times 10^{-2}$ (all in lattice units).}
\label{table:D}
\end{table}
\addtolength{\tabcolsep}{-6pt}

As shown earlier, the solutions agree very well for $\tau = 1$. However, the error may be orders of magnitude larger for $\tau \neq 1$. Depending on the particular application, we may require ratios of diffusion constants that vary significantly (such as in multi-layer systems); however, the errors indicated here may cause the numerical method to appear less than ideal. We discuss the lower asymptotic limit $\tau \to 0.5$ later.

We therefore wish to examine the origin and nature of the $\tau$-dependent error. Of note is that the derivation of the lattice diffusion equation given above (and used heavily in the literature) is done with only a second-order Taylor approximation. To determine the degree to which this approximation leads to the errors shown, we next perform a fourth-order correction to this diffusion equation.

\section{Fourth-order limit of diffusion equation} \label{sec:4th}
In order to introduce a correction to the diffusion equation, we perform a Taylor expansion of the lattice Eqn. (\ref{eqn:lb}) to account for higher orders. As shown by Wagner \cite{PhysRevE.74.056703}, this equation expanded to the fourth order takes the form 
\begin{multline}
(\partial_t + v_{i\alpha}\nabla_\alpha)f_i^0 - \left(\tau - \frac{1}{2}\right)(\partial_t + v_{i\alpha}\nabla_\alpha)^2 f_i^0 + \left(\tau^2 -\tau + \frac{1}{6}\right)
(\partial_t + v_{i\alpha}\nabla_\alpha)^3 f_i^0 \\
- \left(\tau^3 - \frac{3}{2}\tau^2 + \frac{7}{12} \tau -\frac{1}{24}\right)(\partial_t + v_{i\alpha}\nabla_\alpha)^4 f_i^0 
+ O(\partial^5)= \frac{1}{\tau}(f_i^0 - f_i).
\label{eqn:expanded}
\end{multline}

Since we have now introduced higher-order powers into this expansion, we must utilize moments up to the fourth-order. Using the form of the equilibrium distribution in Eqn. (\ref{eqn:eqdist}), we calculate the higher-order moments:
\begin{align}
\sum_i f_i^0 &= \rho\\
\sum_i v_{i\alpha}f_i^0 &= 0\\
\sum_i v_{i\alpha}v_{i\beta}f_i^0 &= \rho\theta\delta_{\alpha\beta}\\
\sum_i v_{i\alpha}v_{i\beta}v_{i\gamma}f_i^0 &= 0\\
\sum_i v_{i\alpha}v_{i\beta}v_{i\gamma}v_{i\delta}f_i^0 &= \frac{\rho\theta}{3}(\delta_{\alpha\beta}\delta_{\gamma\delta} + \delta_{\alpha\gamma}\delta_{\beta\delta} + \delta_{\alpha\delta}\delta_{\beta\gamma})
\end{align}

Summing over all indices of Eqn. (\ref{eqn:expanded}) using these revised moments, we are left with 
\begin{multline}
\partial_t\rho - A(\tau)(\partial_t^2\rho + \nabla_\alpha\nabla_\beta\rho\theta\delta_{\alpha\beta}) + \\
B(\tau)(\partial_t^3\rho + \partial_t\nabla_\alpha\nabla_\beta\rho\theta\delta_{\alpha\beta} + \partial_t\nabla_\alpha\nabla_\gamma\rho\theta\delta_{\alpha\gamma} + \partial_t\nabla_\beta\nabla_\gamma\rho\theta\delta_{\beta\gamma}) - \\
C(\tau)\bigg(\partial_t^4\rho + \partial_t^2\nabla_\alpha\nabla_\beta\rho\theta\delta_{\alpha\beta} + \partial_t^2\nabla_\alpha\nabla_\gamma\rho\theta\delta_{\alpha\gamma} + \partial_t^2\nabla_\alpha\nabla_\delta\rho\theta\delta_{\alpha\delta} + \partial_t^2\nabla_\beta\nabla_\gamma\rho\theta\delta_{\beta\gamma} \\
+ \partial_t^2\nabla_\beta\nabla_\delta\rho\theta\delta_{\beta\delta} + \partial_t^2\nabla_\gamma\nabla_\delta\rho\theta\delta_{\gamma\delta} + \nabla_\alpha\nabla_\beta\nabla_\gamma\nabla_\delta \left[\frac{\rho\theta}{3}(\delta_{\alpha\beta}\delta_{\gamma\delta} + \delta_{\alpha\gamma}\delta_{\beta\delta} + \delta_{\alpha\delta}\delta_{\beta\gamma})\right]\bigg) + O(\partial^5) = 0
\label{eqn:uglysum}
\end{multline}
where we have defined the $\tau$-dependent prefactors
\begin{eqnarray*}
A(\tau) &\equiv& \tau - \frac{1}{2} \\
B(\tau) &\equiv& \tau^2 - \tau + \frac{1}{6} \\
C(\tau) &\equiv& \tau^3 - \frac{3}{2}\tau^2 + \frac{7}{12}\tau - \frac{1}{24}
\end{eqnarray*}
for brevity.

\begin{figure}
    \centering
    \includegraphics[width=0.5\columnwidth,clip=true]{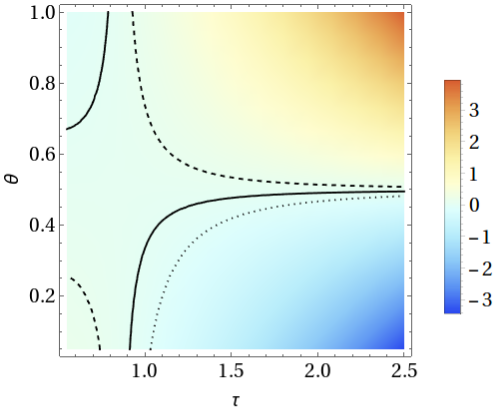}
    \caption{Density field representation of $\alpha(\tau,\theta)/D(\tau,\theta)$, with contour lines indicating $\alpha(\tau,\theta)/D(\tau,\theta) = 0$ (solid), $-1/\pi^2$ (dotted, predicted instability), $1/\pi^2$ (dashed, shown for symmetry).}
    \label{fig:alphaD}
\end{figure}

This form is not particularly useful since there are mixed spatial and temporal derivatives in the higher-order powers. In our one-dimensional implementation, we can drop our indices. We use the diffusion equation to write the temporal derivatives in terms of the spatial derivatives as
\begin{equation}
\partial_t\rho = \left(\tau - \frac{1}{2}\right)\nabla_\alpha^2 \rho\theta + O(\nabla^3).
\end{equation}
It immediately follows that 
\begin{equation}
\partial_t^2 \rho = \left(\tau - \frac{1}{2}\right)^2\nabla_\alpha^2\nabla_\beta^2\rho\theta^2 +O(\nabla^5).
\end{equation}
We can then introduce these two substitutions into Eqn. (\ref{eqn:uglysum}) and we have 
\begin{multline}
\partial_t\rho - \left(\tau - \frac{1}{2}\right)\nabla^2\rho\theta - \left(\tau - \frac{1}{2}\right)^3\nabla^4\rho\theta^2 + \left(\tau^2 -\tau + \frac{1}{6}\right) 
\left(\tau-\frac{1}{2}\right)3\nabla^4\rho\theta^2 \\
- \left(\tau^3 -\frac{3}{2}\tau^2 + \frac{7}{12}\tau - \frac{1}{24}\right)\nabla^4\rho\theta = 0.
\end{multline}
We then obtain the form of a corrected diffusion equation
\begin{equation}
\partial_t\rho =D\nabla^2\rho + \alpha\nabla^4\rho
\label{eqn:correctedfull}
\end{equation}
with corrections up to the fourth power in spatial derivatives, where we define
\begin{equation}
\alpha = \alpha(\tau,\theta) \equiv \left(2\tau^3\theta - \tau^3 -3\tau^2\theta + \frac{3}{2}\tau^2 + \right. \\
\left. \frac{5}{4}\tau\theta -\frac{7}{12}\tau - \frac{1}{8}\theta + \frac{1}{24}\right)\theta.
\label{eqn:alpha}
\end{equation}
This definition of $\alpha$ represents the expected error between the second-order diffusion equation and the corrected fourth-order equation. For certain parameter values, such as $\tau = 1$ and $\theta = 1/3$, we have $\alpha = 0$, which accounts for higher accuracy observed for such parameters. We plot a density field representation of the relative error quantity $\alpha(\tau,\theta)/D(\tau\theta)$ in Fig. \ref{fig:alphaD}. We indicate a contour where this quantity vanishes, as well as additional contours whose numerical importance will be explained in later sections.

The correction term has a similar form to a surface tension term in a Cahn-Hilliard equation. In this case, positive values of $\alpha$ would correspond to a negative surface free energy. This implies that simulations with positive $\alpha$ should be unstable for high frequency perturbations. This equation can be solved in Fourier space, allowing us to verify our analytical predictions with lattice Boltzmann simulations. In the subsequent section, we perform this analysis.

\section{Fourier analysis of correction term} \label{sec:fourier}
A Fourier transform of Eqn. (\ref{eqn:correctedfull}) yields
\begin{equation}
\partial_t \widehat{\rho}(k,t,\alpha) = -Dk^2\widehat{\rho}(k,t,\alpha) -\alpha k^4\widehat{\rho}(k,t,\alpha).
\label{eqn:FTdiffusion}
\end{equation}
Here $k$ is any specific Fourier mode and $\widehat{\rho}(k,t)$ is the $k$-space density represented by
\begin{equation}
\widehat{\rho}(k) = \frac{1}{2\pi}\int_0^{L_x} \rho(x) e^{\frac{2\pi ikx}{L_x}}dx,
\label{eqn:integral}
\end{equation}
where $L_x$ is the system size in the $x$-direction. Even though $x$ is continuous, the finite periodicity of $2\pi$ causes $k$ to be discrete. This allows for our system to contain a finite number of $k$ modes which can be now examined independently. The form of Eqn. (\ref{eqn:FTdiffusion}) is simple since different $k$ modes do not couple. In $k$-space, the initial profile at $t=0$ is chosen by defining $\rho(x,0)$, which for $\widehat{\rho}(k,0)$ gives Eqn. (\ref{eqn:integral}) and
\begin{equation}
\widehat{\rho}(k,t,\alpha) = \widehat{\rho}(k,0) e^{-(Dk^2t + \alpha k^4t)}.
\end{equation}
We reproduce the uncorrected diffusion equation by setting $\alpha = 0$, obtaining
\begin{equation}
\widehat{\rho}(k,t,0) = \widehat{\rho}(k,0) e^{-Dk^2t}.
\end{equation}
These predictions are implemented on a discrete lattice which implies that there will be a finite number of $k$ modes. From Eqn. (\ref{eqn:integral}), we have
\begin{equation}
k=\frac{2\pi}{L_x}
\label{eqn:kdef}
\end{equation}
which implies a maximum allowed $k$ mode when $k=\pi$ and a minimum lattice dimension of $L_x = 2$. In this finite system, we have the back transform
\begin{equation}
\rho(x,t) = \sum_k e^{ikx}\widehat{\rho}(k,t,\alpha).
\label{eqn:backxform}
\end{equation}

It is now possible to verify this theoretical prediction by examining the decay of specific Fourier modes by imposing an initial profile
\begin{equation}
\rho(x,0) = \sin(k x).
\end{equation}
Using this profile, the uncorrected and corrected $k$-space densities become, respectively,
\begin{align}
\widehat{\rho}(k,t,0) =& \sin(k x) e^{-tDk^2}\nonumber\\
\widehat{\rho}(k,t,\alpha) =& \sin(k x) e^{-t(Dk^2 + \alpha k^4)}.
\label{eqn:kspacerhos}
\end{align}
In practice, we change $k$ by varying the system size $L_x$. An interesting point to note is that when $\alpha < -\frac{D}{\pi^2}$, it is predicted that the numerical simulations would be unstable. This is predicted due to the fact that in Eqn. (\ref{eqn:kspacerhos}), the negative $\alpha$ term leads to a positive exponent and causes $\widehat{\rho}(k,t,\alpha)$ not to decay.

\section{Numerical verification of correction term} \label{sec:numerical}
To determine the validity of the prediction for the correction term shown in Eqn. (\ref{eqn:alpha}), we define a ratio between the two forms of k-space density in Eqn. (\ref{eqn:kspacerhos}) in a simple form such that
\begin{equation}
R(k,t,\alpha) \equiv \frac{\widehat{\rho}(k,t,0)}{\widehat{\rho}(k,t,\alpha)} = e^{\alpha k^4 t}. \label{eqn:Rratio}
\end{equation}

\begin{figure}
    \includegraphics[width=0.5\columnwidth,clip=true]{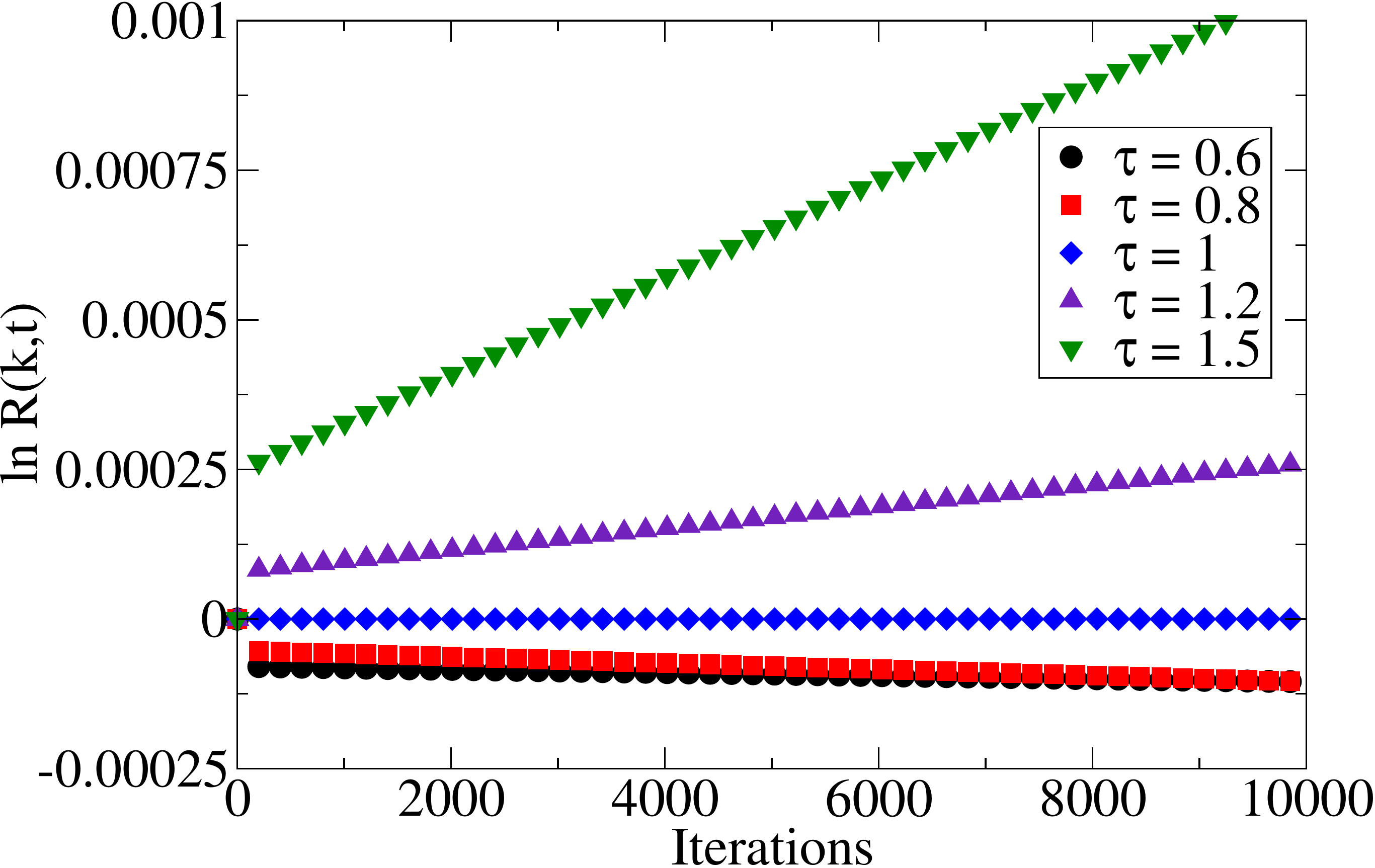}
    \caption{A plot of $\ln R(k,t)$ as a function of discrete time steps for various values of $\tau$ and $\theta=1/3$ and $L_x = 200$. It is observed that there is an initial offset in $\ln R(k,t)$. As the system evolves, we see that the behavior does decay as expected. Since there is this initial offset, we cannot use these early times when calculating the derivative in Eqn. (\ref{eqn:rk}).}
    \label{fig:rktime}
\end{figure}
We can use this relation to measure $\alpha$ from numerical simulations. We do this by initializing our probability distributions by $f_i(x,0) = f_i^0(\sin(2\pi x/L_x))$ and then varying $L_x$. Our first prediction is that $\ln(R(k,t))$ is a linear function of $t$. We can find $\alpha$ from the time evolution of the density through
\begin{equation}
\alpha_{\operatorname{exp}} = \frac{1}{k^4}\frac{d}{dt}\ln R(k,t)
\label{eqn:rk}
\end{equation}
where we numerically calculate the temporal derivative using a finite difference method.

\begin{figure}
 \includegraphics[width=0.5\columnwidth,clip=true]{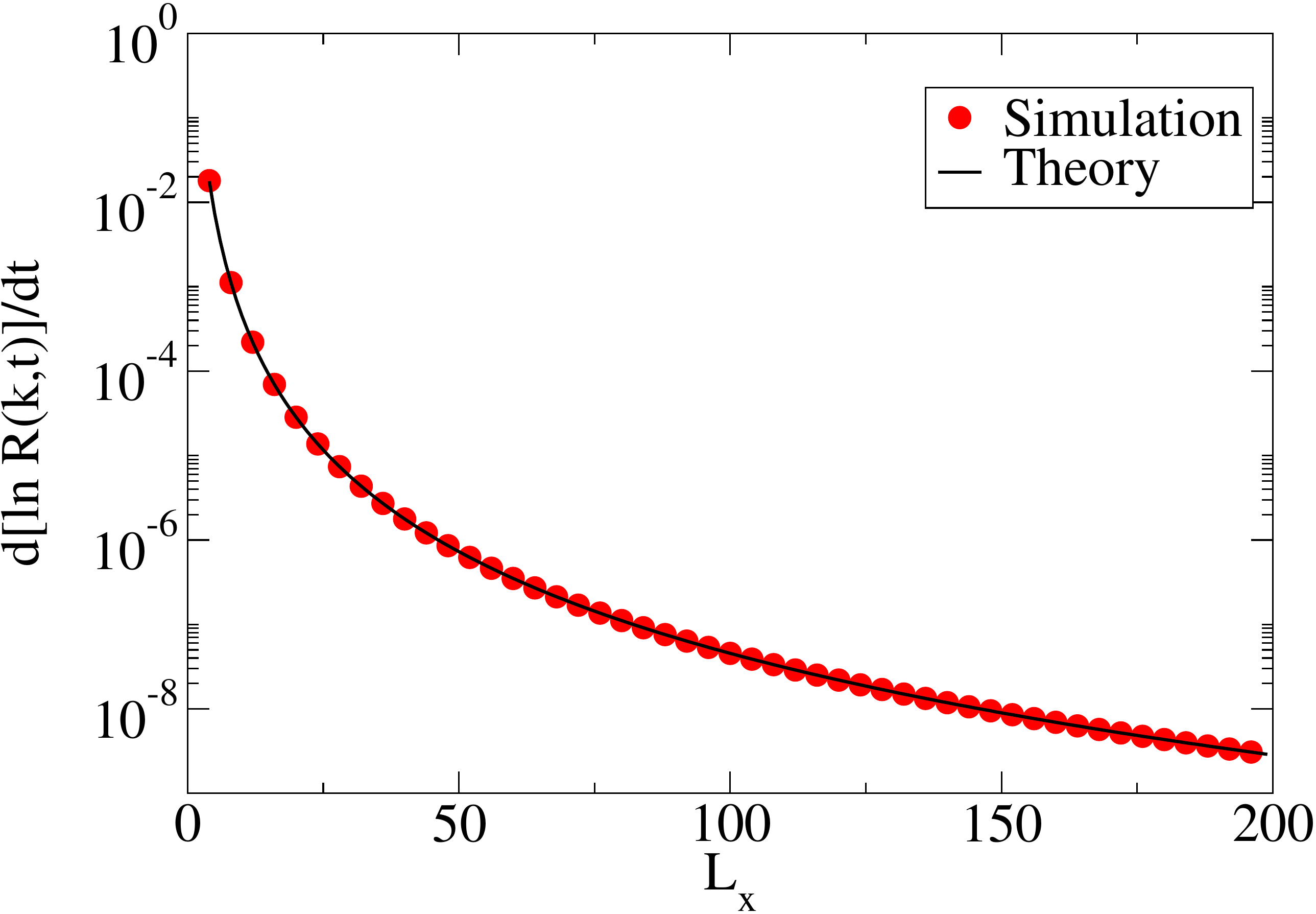}
  \caption{Logarithmic representation of $\frac{d}{dt}\left[\ln R(k,t)\right]$ as a function of $k$ from simulation data for $\tau = 1$ and $\theta = 0.1$. Good agreement is observed between the simulation and the curve fit for up to $L_x=200$.}
  \label{fig:rktau1}
\end{figure}

The numerical evaluation of Eqn. (\ref{eqn:Rratio}) using the numerical results is shown in Fig. \ref{fig:rktime}. At $t=0$ we have $R=1$ by construction, but for all $\tau \neq 1$ we observe a rapid transient change which manifests itself as a near instantaneous jump in Fig. \ref{fig:rktime}. After this transient period, the behavior of $\ln(R)$ is indeed linear, as expected.  We then calculate
\begin{equation}
\frac{d}{dt}\ln R(k,t) \approx \frac{\ln R(k,t_2) - \ln R(k,t_1)}{t_2 - t_1},
\end{equation}
where we take $t_1$ when $\widehat{\rho}(k,t,\alpha) = 0.5$ and $t_2$ when $\widehat{\rho}(k,t,\alpha) = 0.01$ to avoid any difficulties with the offset. Eqn. (\ref{eqn:rk}) gives our correction polynomial as a function of any Fourier mode $k$. Using this form, we can compare our predicted correction term in Eqn. (\ref{eqn:alpha}) to a numerical representation. Fig. \ref{fig:rktau1} shows simulation data for $\frac{d}{dt}\left[\ln R(k,t)\right]$ for $\tau = 1$ and $\theta = 0.1$. We see a good fit for all $k$ modes between simulation and the prediction in Eqn. (\ref{eqn:rk}).

\begin{figure}
  \includegraphics[width=0.5\columnwidth,clip=true]{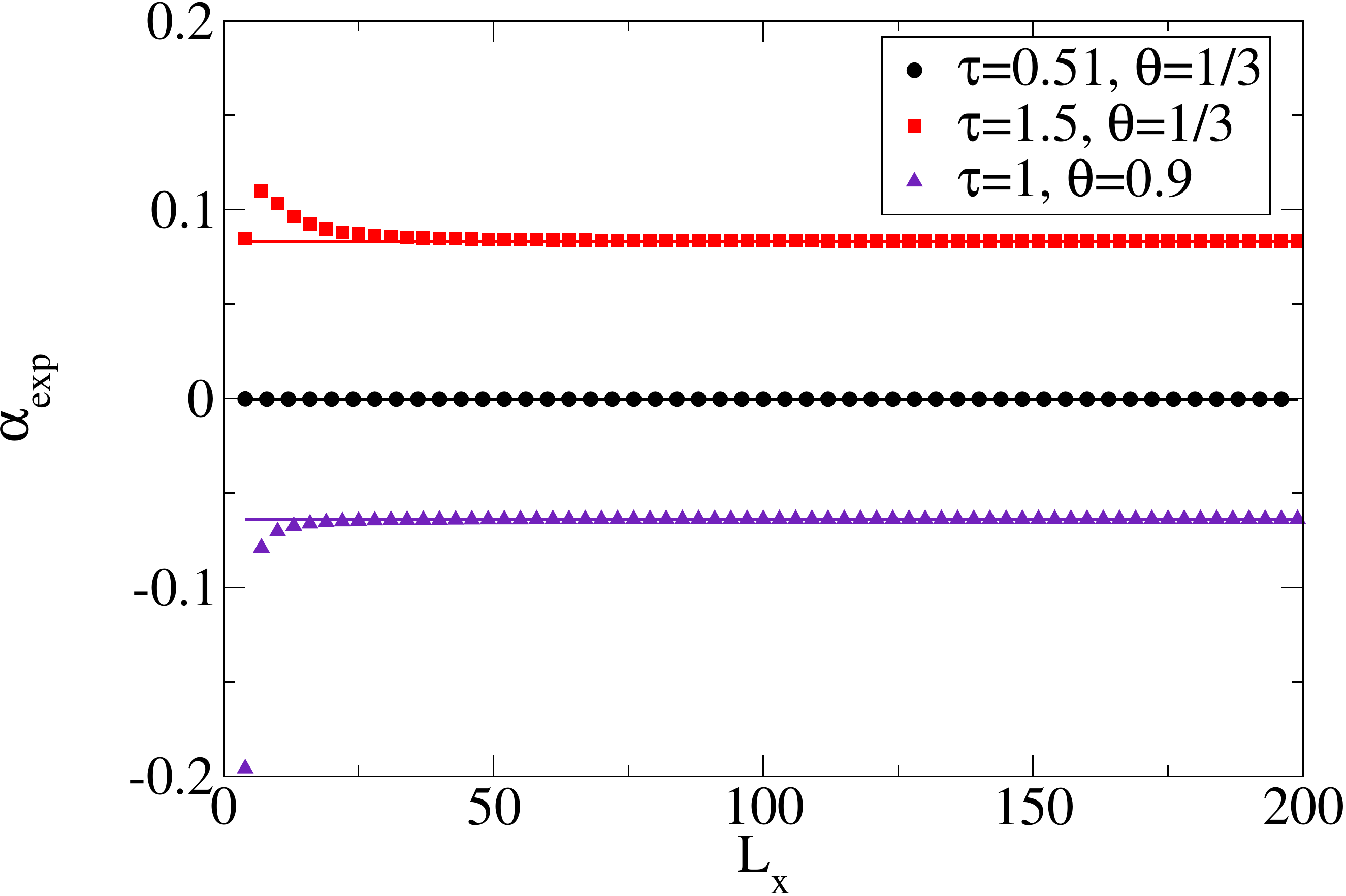}
  \caption{Comparison of $\alpha_{\operatorname{exp}}$ (symbol) to theoretical prediction for $\alpha$ from Eqn. (\ref{eqn:alpha}) (solid line) for various values of $\theta$ and $\tau$ as a function of $L_x$. It is observed that for $\tau=0.51$ and $\theta = 1/3$ that $\alpha_{\operatorname{exp}}$ matches the theoretical $\alpha$ well for all $L_x$. For sets of values $\tau=1$ with $\theta = 0.9$  and $\tau=1.5$ with $\theta = 1/3$, there is a good match for $L_x > 40$ but deviations are observed for small values of $L_x$.}
  \label{fig:k4tau15}
\end{figure}

We first test the prediction comparing $\alpha_{\operatorname{exp}}$ in Eqn. (\ref{eqn:rk}) to our theoretical prediction for $\alpha$ from Eqn. (\ref{eqn:alpha}). Fig. \ref{fig:k4tau15} shows a comparison between $\alpha_{exp}$ and our theoretical prediction for $\alpha$ for various values of $\tau$ and $\theta$ as a function of $L_x$. For this analysis, we chose a known stable value for either $\tau$ or $\theta$ and set the other parameter as a more extreme value. For a choice of $\theta=1/3$, we set $\tau=0.51$ as the extreme value. In these cases, we see very good agreement between $\alpha_{\operatorname{exp}}$ and our prediction. In the cases of $\theta=1/3$ with $\tau = 1.5$ and $\theta=0.9$ and $\tau = 1$ we observe good agreement for $L_x > 40$, but as $L_x$ becomes smaller, deviations begin to increase. This suggests that there is a discrepancy in $\alpha_{\operatorname{exp}}$ for large $k$ modes. 

In the case where $\alpha = 0$, it is interesting to note that the results match a $\frac{1}{k^6}$ rather than the predicted $\frac{1}{k^4}$ fit. This implies that there are additional correction terms which may be relevant at specific values of $\tau$ and $\theta$. These higher-order corrections are not considered in the present analysis.

As discussed previously, Eqn. (\ref{eqn:kspacerhos}) predicts numerical instability when $\alpha < -\frac{D}{\pi^2}$. The density representation shown in Fig. \ref{fig:alphaD} implies that this will happen as we increase $\tau$ and decrease $\theta$ to extreme values ($\tau \gtrsim 4$ and $\theta \lesssim 0.3$ simultaneously). A contour showing $\alpha(\tau,\theta)/D(\tau,\theta) = -1/\pi^2$, the start of the region of instability, is shown in that figure.

\begin{figure}
    \includegraphics[width=0.5\columnwidth,clip=true]{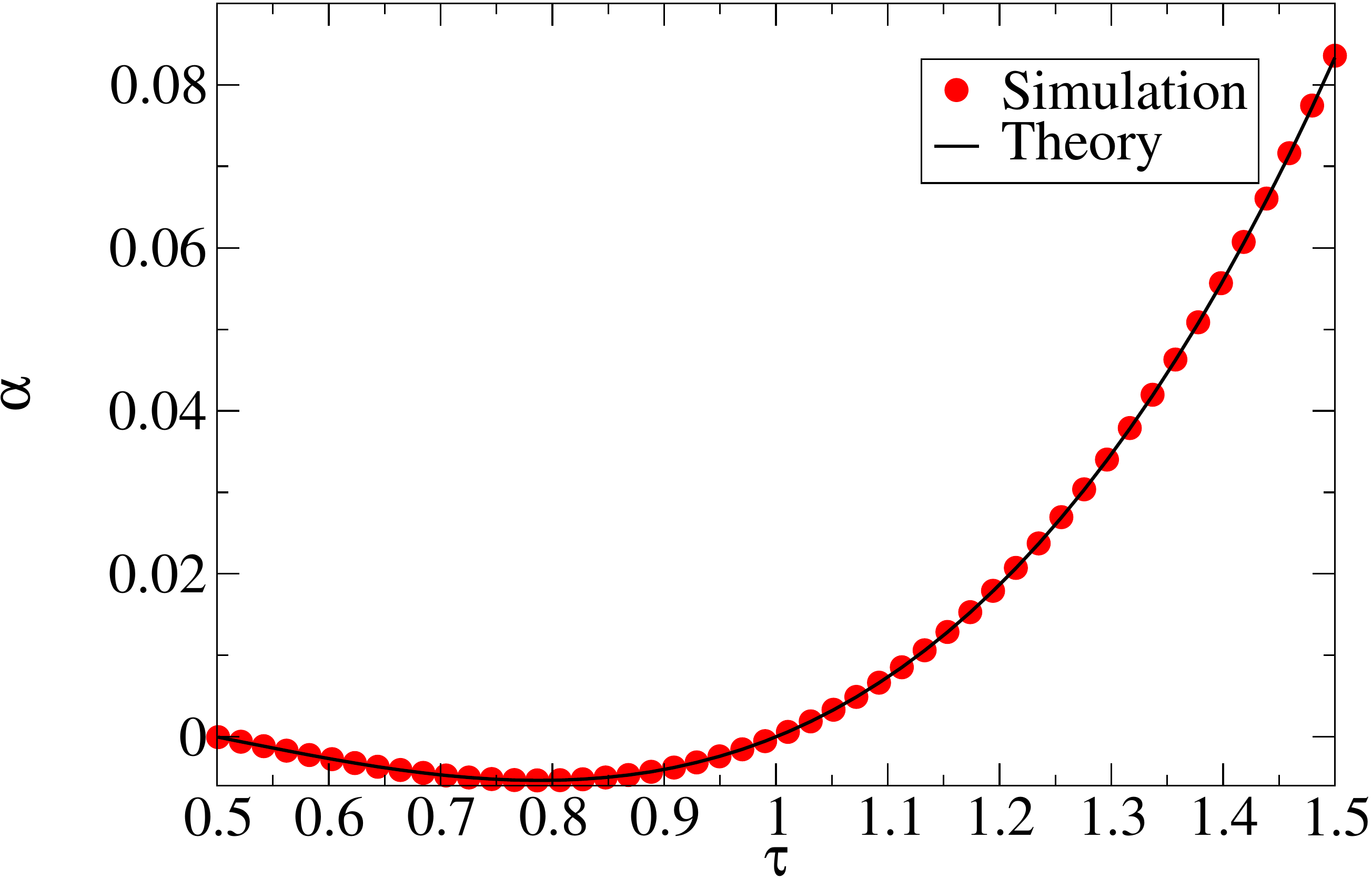}
    \caption{Comparison of numerical results and theoretical $\alpha$ as a function of $\tau$, with $\theta = \frac{1}{3}$. Results are collected for $L_x=100$.}
    \label{fig:alphatau}
\end{figure}

\begin{figure}
    \includegraphics[width=0.5\columnwidth,clip=true]{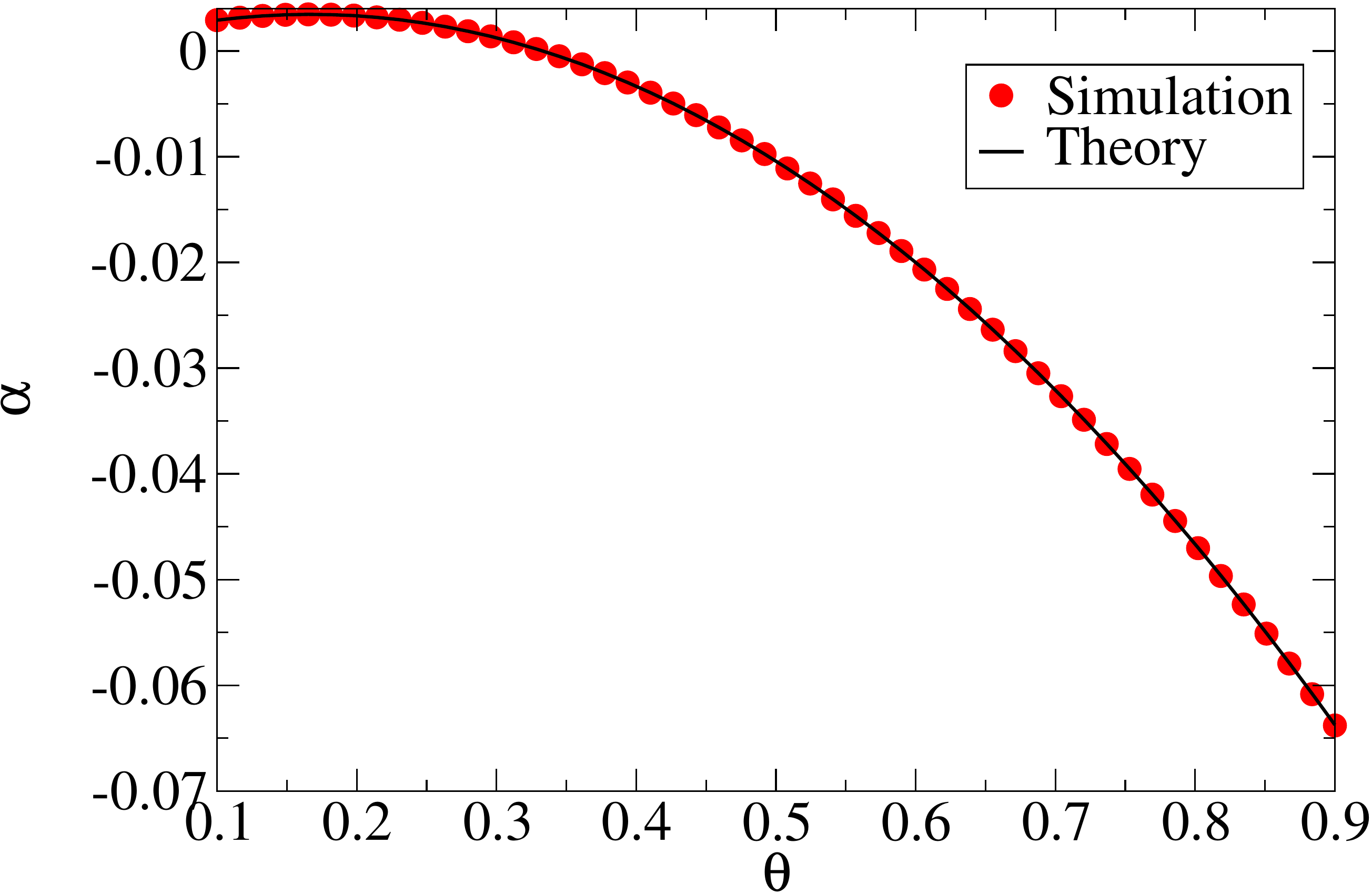}
    \caption{Comparison of numerical results and theoretical $\alpha$ as a function of $\theta$, with $\tau = 1$. Results are collected over 100 independent $k$ modes.}
    \label{fig:alphatheta}
\end{figure}

It is instructive to examine $\alpha$ while holding either $\tau$ or $\theta$ fixed. Setting $\theta=\frac{1}{3}$, we examine $\alpha$ as a function of $\tau$ alone in Fig. \ref{fig:alphatau}, which shows excellent agreement to theory over 100 independent $k$ modes. We set $\tau=1$ and examine $\alpha$ as a function of $\theta$ alone in Fig. \ref{fig:alphatheta}, with similarly excellent agreement.

\section{Application of correction to reservoir diffusion} \label{sec:application}
\begin{figure}
    \centering
    \includegraphics[width=0.5\columnwidth,clip=true]{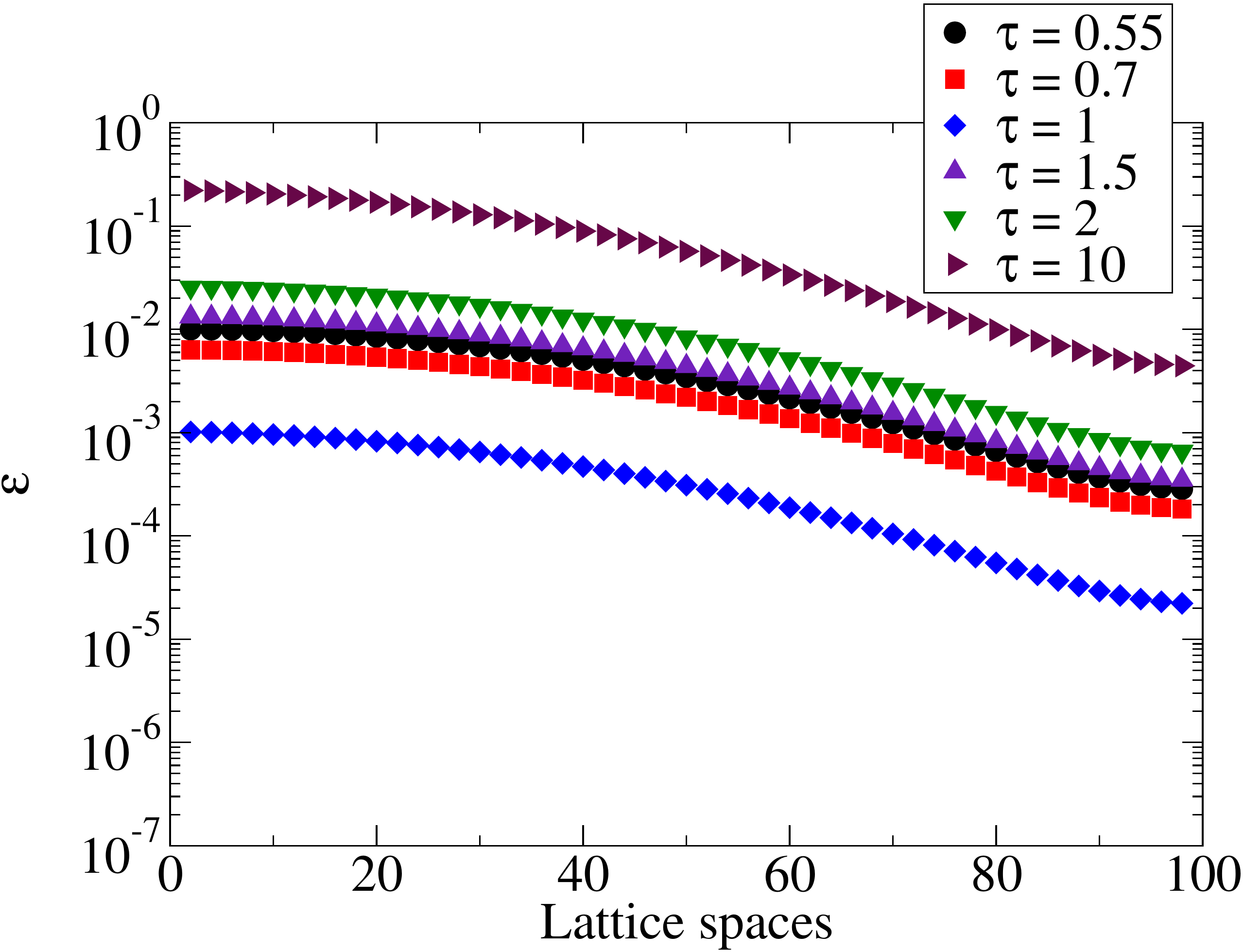}
    \caption{Absolute error profile $\epsilon$ between numerical and fourth-order Fourier analytical concentration at various $\tau$. All simulations were run to the same scaled time, corresponding to four hours.}
    \label{fig:error-finite-alpha}
\end{figure}

With the fourth-order correction term in hand and its correctness assured, we next determine its applicability to our reservoir coating system. Fig. \ref{fig:error-finite-alpha} shows the absolute error profile between lattice Boltzmann simulation results and a fourth-order corrected analytical solution. This solution is produced by first setting up an appropriate initial step function
\begin{equation}
\rho(x,0) = \left\{
\begin{array}{lll}
2 & , & L_x < x < 3L_x \\
1 & , & x = L_x \text{ or } x = 3L_x \\
0 & , & \text{else}
\end{array}\right.
\label{eqn:periodic}
\end{equation}
in a periodic lattice. This is entirely equivalent to the boundary conditions implied by the derivation of the second-order error function solution in Eqn. (\ref{eqn:rho}). We transform this step function into $k$ space via a discrete Fourier transform, use the fourth-order correction to perform a time evolution, and then transform the result back into real space. Strictly speaking, the method of Eqn. (\ref{eqn:rho}) generates a continuous solution, while the Fourier transform approach yields a discrete solution. We discuss the ramifications of this difference in the Appendix and conclude that the difference in solution discretization is very small and of the same order as the error produced in our best numerical results.

We note no consistent improvement over the second-order shown in Fig. \ref{fig:error-finite} from the introduction of the fourth-order correction. However, the magnitude of the error, especially for high values of $\tau$, remains troubling. The nature of the finite simulation lattice is such that the boundaries are treated independently of other lattice sites. In particular, the reservoir density $\rho$ is set manually and not strictly determined by local distributions. Since we have seen that $\tau$-dependent errors tend to accumulate near the reservoir boundary over an order of magnitude higher than at the substrate boundary, the nature of using such a finite lattice is suspect. The case when $\tau=1$ yielded excellent agreement throughout the finite lattice, but this is consistent with the immediate relaxation of local equilibrium distributions and does not apply to other values of $\tau$.

This $\tau$-dependent error is consistent with the jump observed in Fig. \ref{fig:rktime}, where setting $f_i(x,0)=f_i^0(\rho(x))$ led to deviations. Indeed, Eqn. (\ref{eqn:expanded}) implies that 
\begin{equation}
f_i=f_i^0-\tau (\partial_t f_i^0(\rho)+v_{i\alpha} \nabla_\alpha f_i^0(\rho))+O(\partial^2),
\end{equation}
which suggests an approach that would allow us to increase the accuracy of our boundary conditions.

\begin{figure}
    \centering
    \includegraphics[width=0.4\columnwidth,clip=true]{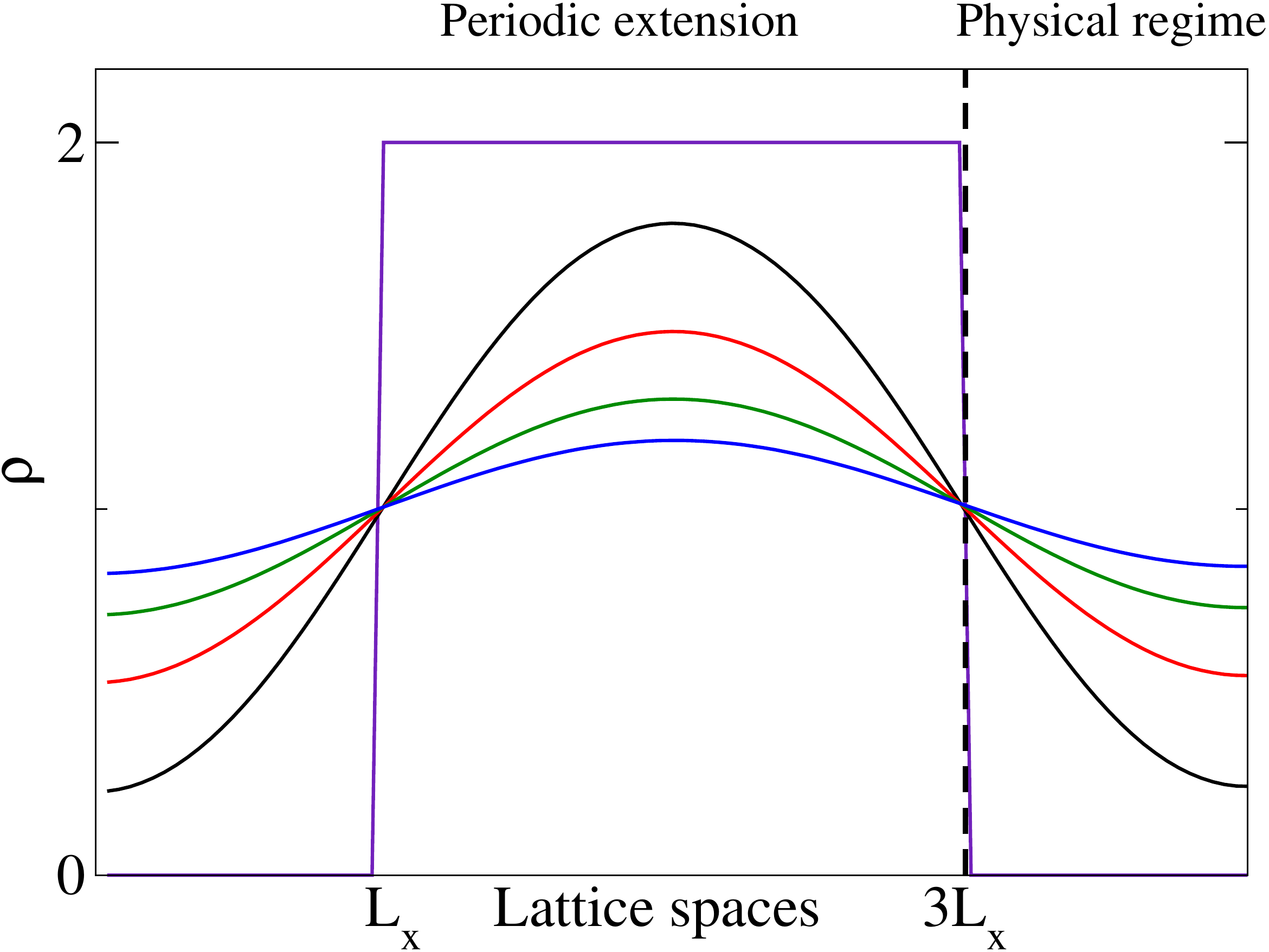}
    \caption{Periodic step function from Eqn. \ref{eqn:periodic}, with reservoir at $x = 3L_x$ and substrate at periodic boundary $x = 4L_x$.}
    \label{fig:periodic}
\end{figure}

In our current case, however, we can avoid the cumbersome issue of the boundary condition altogether by simply embedding the system into the periodic lattice used for establishing the initial step function condition of the analytical Fourier solution. This permits a more standard lattice Boltzmann approach that does not rely on manual density adjustment at the reservoir (here at $x = 3L_x$) and uses symmetry to establish the substrate (at $x = 4L_x$) with no bounceback. With this setup, we have removed the need for boundary conditions altogether; we therefore expect that the $\tau$-dependent error should be substantially reduced, especially at the reservoir boundary. Fig. \ref{fig:periodic} shows a diagram of the periodic step function from Eqn. \ref{eqn:periodic} used for this analysis.

\begin{figure}
    \centering
    \includegraphics[width=0.5\columnwidth,clip=true]{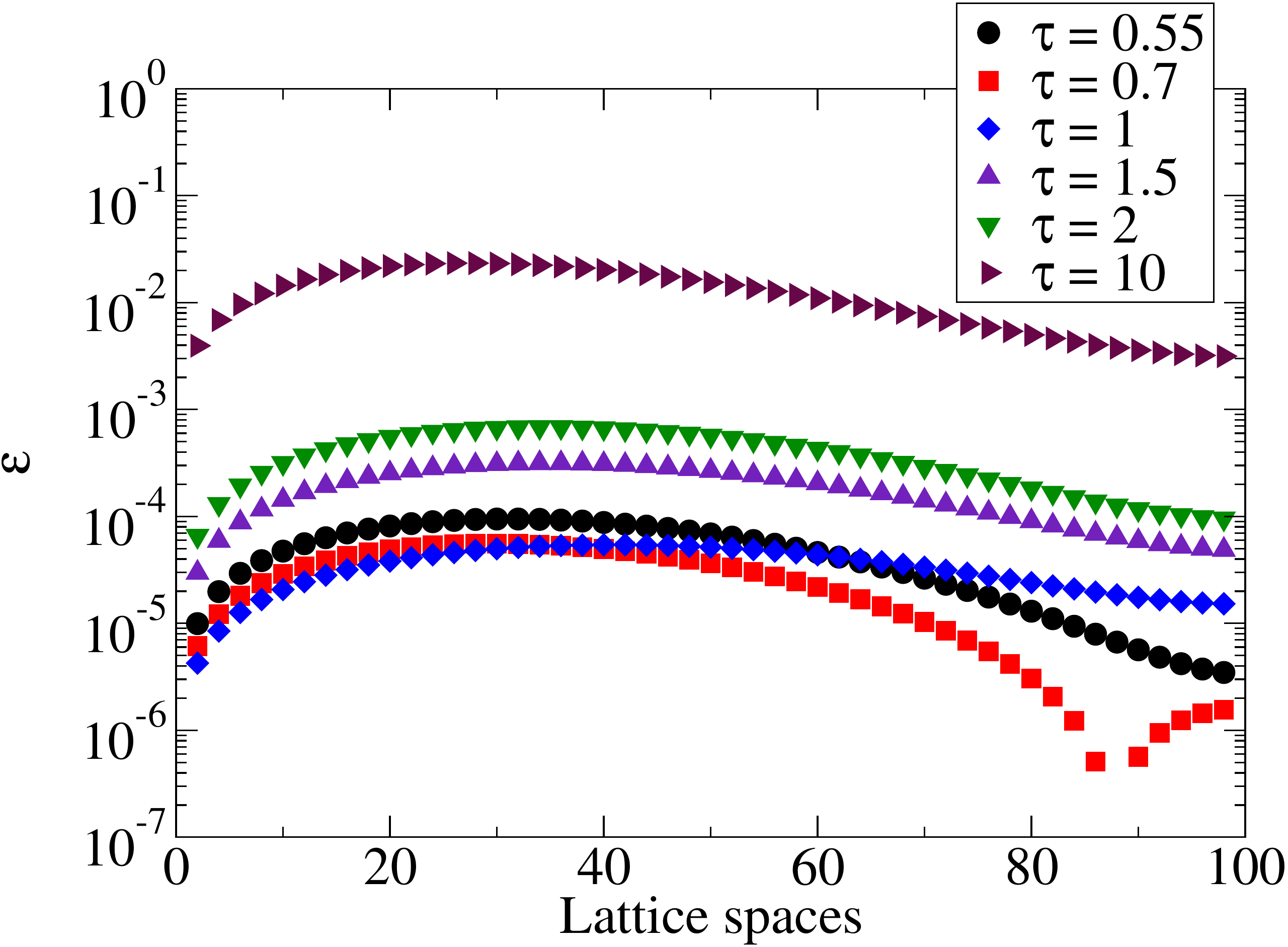}
    \caption{Periodic system absolute error profile $\epsilon$ between numerical and second-order Fourier analytical concentration at various $\tau$. All simulations were run to the same scaled time, corresponding to four hours.}
    \label{fig:error-periodic}
\end{figure}

\begin{figure}
    \centering
    \includegraphics[width=0.5\columnwidth,clip=true]{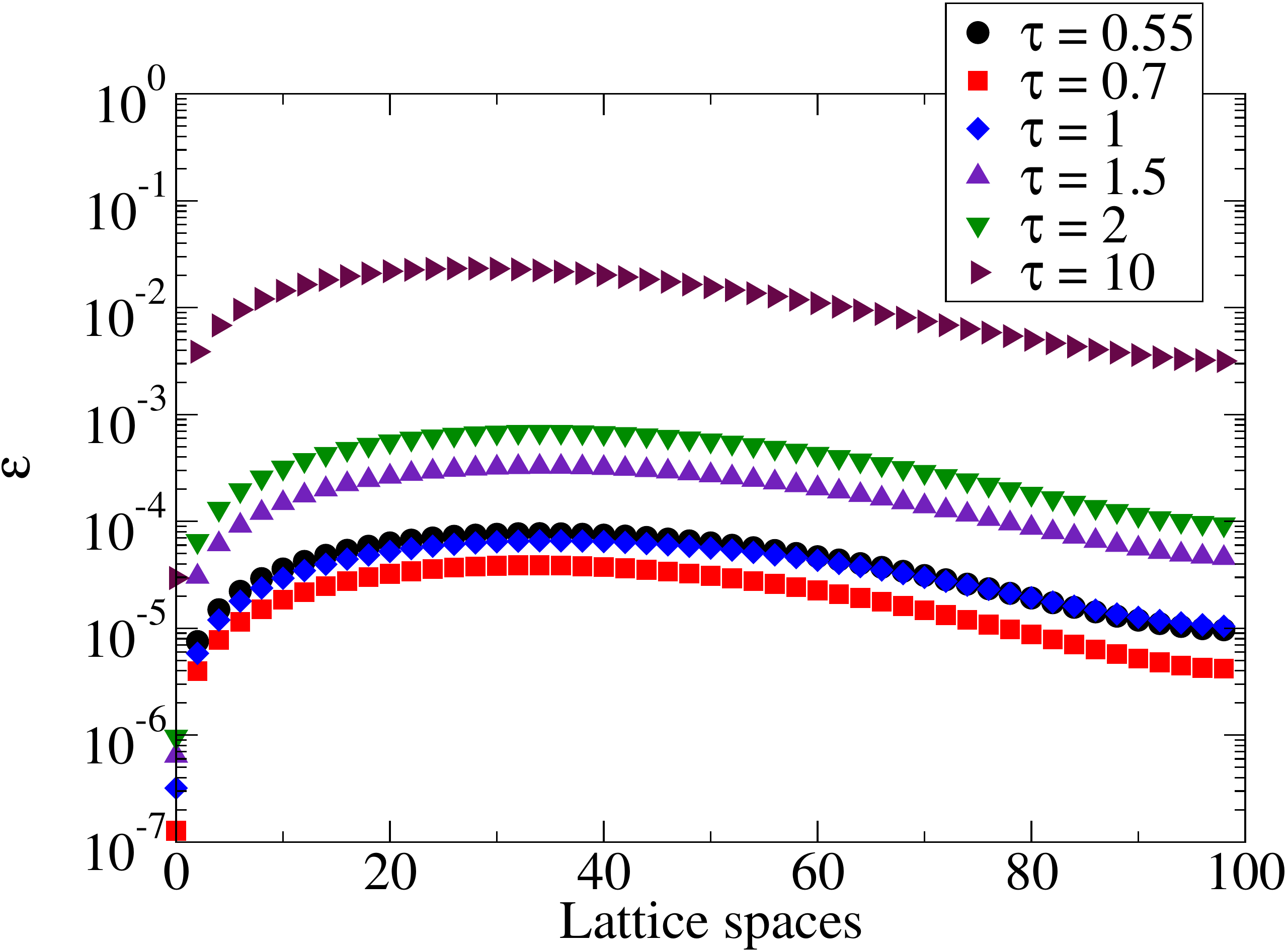}
    \caption{Periodic system absolute error profile $\epsilon$ between numerical and fourth-order Fourier analytical concentration at various $\tau$. All simulations were run to the same scaled time, corresponding to four hours.}
    \label{fig:error-periodic-alpha}
\end{figure}

We again run two sets of simulations for our range of $\tau$ values to the same scaled time, both using the periodically-embedded lattice simulation. The first set of simulations uses only the traditional second-order approximation and is shown in Fig. \ref{fig:error-periodic}. The second set applies our fourth-order correction and is shown in Fig. \ref{fig:error-periodic-alpha}.

As was hoped, the error at the reservoir is reduced by orders of magnitude when compared to the finite system with imposed boundaries. This confirms that the accumulated error from the finite system is due to the presence of boundary conditions that are only guaranteed to match at $\tau = 1$ when relaxation is immediate during collisions. However, contrary to expectation, there is almost no benefit from the fourth-order $\alpha(\tau,\theta)$ correction, even though its validity was verified via Fourier analysis.

It is of value to mention here the effects of using a progressively lower of $\tau$, since our results suggest that errors increase with $\tau$, regardless of the choice of boundary conditions considered. Of course, a lower choice of this parameter leads to increasingly long simulations, which must be balanced with the desired numerical accuracy. We examined a range of $\tau$ values as low as $\tau = 0.5001$ and found essentially no change in error from the $\tau = 0.55$ lower limit presented throughout this paper. This suggests that (to within machine accuracy), there is likely no theoretical limit to the ratio of diffusion constants possible. This implies that a study of a multi-layer coating stack, which fixes the diffusitivity ratio through a selection of $\tau$ values, is possible for a situation where one coating's diffusion constant is orders of magnitude higher than the other.

It is natural at this point to wonder if there are any choices of parameters $\tau$ and $\theta$ for which the fourth-order correction provides substantial benefit in our reservoir problem, especially since its use in simulations incurs additional computational burden. Naturally, any such error analysis depends heavily on the particular problem of interest, and therefore on the initial profile and desired time evolution. For our system, we examine the parameter space $0.5 < \tau \leq 2.5$ and $0.1 \leq \theta \leq 1.0$. For each point in this space, we run a lattice Boltzmann periodic reservoir system simulation to the same scaled time. After this time, we compute the ratio 
\begin{equation}
\overline{\epsilon} \equiv \frac{\epsilon_4(\tau,\theta)}{\epsilon_2(\tau,\theta)},
\end{equation}
where 
\begin{equation}
\epsilon_{2,4} \equiv \sqrt{\frac{1}{L_x}\sum_{x=1}^{L_x}\left[\rho(x)-\rho_{2,4}(x)\right]^2}
\label{eqn:eps}
\end{equation}
is the root mean square error between numerical concentration $\rho$ and second- or fourth-order Fourier analytical concentration $\rho_{2,4}$. If $\overline{\epsilon} \approx 1$, there is no appreciable correction from using the fourth-order solution; as $\overline{\epsilon} \to 0$, the correction becomes more substantial. From a computational perspective, there is a trade-off between the computational burden of the correction and the benefit (if any) from using it. We do not comment on the appropriate balance for any particular situation.

\begin{figure}
    \centering
    \includegraphics[width=0.5\columnwidth,clip=true]{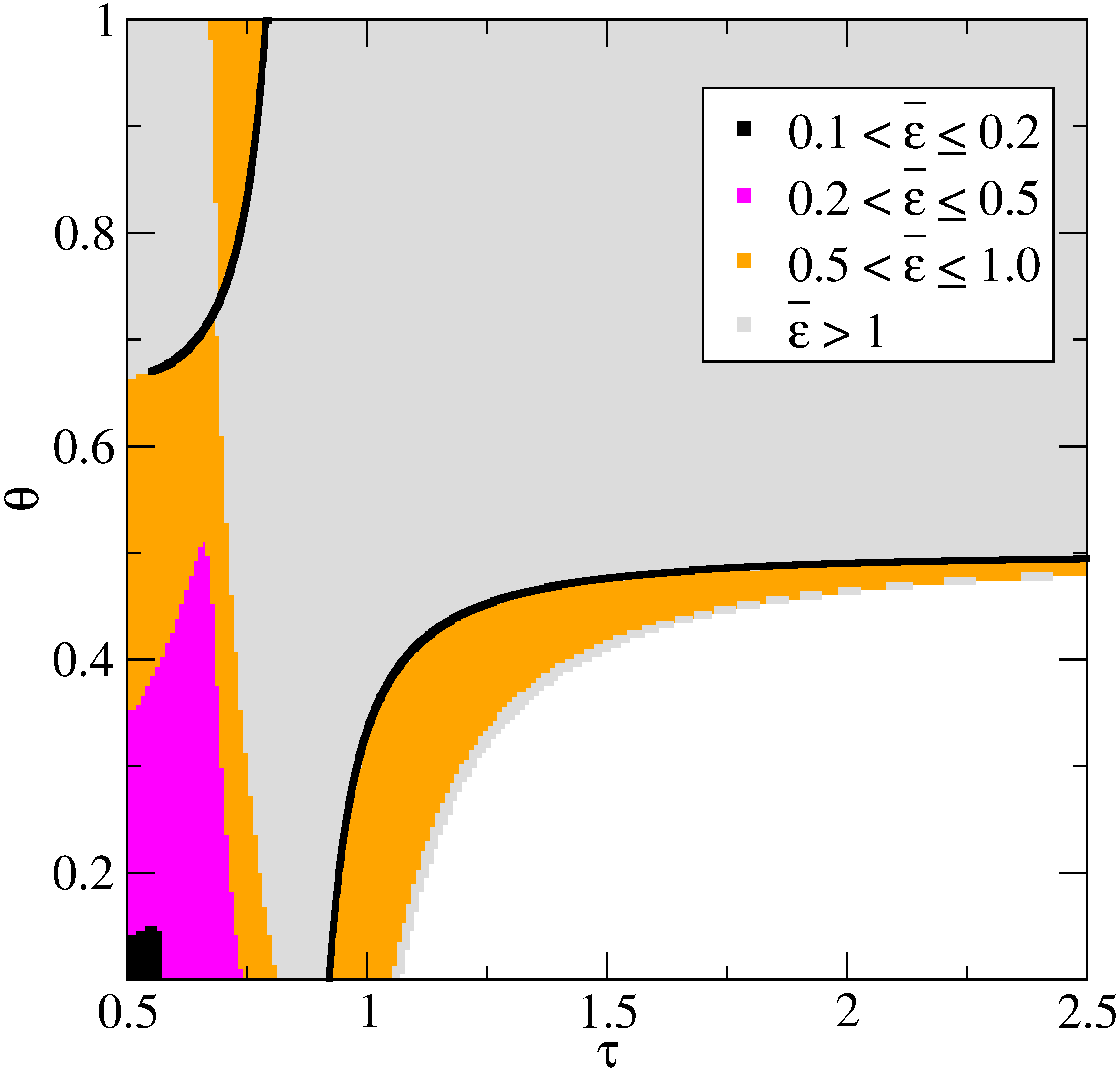}
    \caption{Error ratio $\overline{\epsilon}$, indicating bands comparing the second- and fourth-order Fourier solution accuracy. All simulations were run to the same scaled time, corresponding to a macroscopic system time of $3.5$ seconds. Also shown is the $\alpha(\tau,\theta)=0$ contour (black line).}
    \label{fig:ratio-density}
\end{figure}

\begin{figure}
    \centering
    \includegraphics[width=0.5\columnwidth,clip=true]{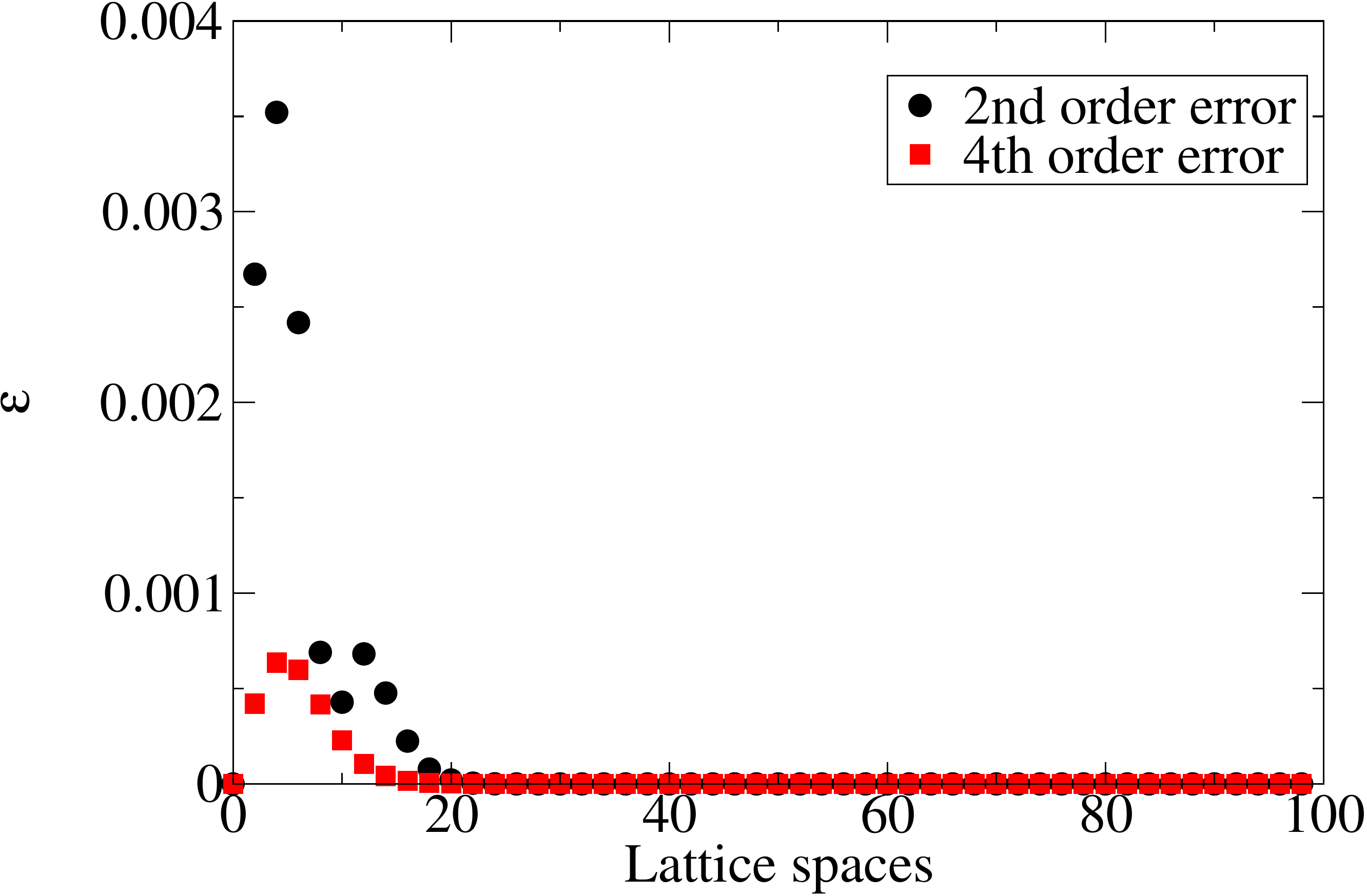}
    \caption{Absolute error between numerical simulation results and second-order Fourier (circles) and fourth-order Fourier (squares) analytical solutions. Simulation was run with $\tau = 0.55$ and $\theta = 0.15$ to the macroscopic equivalent time of $3.5$ seconds.}
    \label{fig:lowepsilon}
\end{figure}

Contrary to expectation, there are no regions of the given parameter space where $\overline{\epsilon} < 0.9$ during long times, indicating no appreciable benefit to the correction. Further, the fourth-order analysis predicts numerical instability in the bulk region of parameter space where $\alpha < -1/\pi^2$, although the numerical simulations and second-order analysis remain stable. This is a surprising result overall: a fourth-order correction is not only unhelpful in increasing the accuracy of solutions at long times, it is often worse than the second-order approximation and predicts numerical problems incorrectly.

If we instead run the same analysis for a much shorter time (in the equivalent macroscopic system, just $3.5$ seconds), the results are more promising and shown in Fig. \ref{fig:ratio-density}. For small values of both $\tau$ and $\theta$, the fourth-order correction increases accuracy by an order of magnitude. This is largely due to the fact that the fourth-order theory accurately predicts some early time oscillations at the sharp reservoir interface, as shown by the error reduction in Fig. \ref{fig:lowepsilon}. This discussion of higher-order effects gives the rather surprising result that for our barrier coating application, there is no noticeable improvement. This may also arise because even the second-order results are accurate enough that any resulting errors are of the same order of magnitude as the difference between continuous and discrete analytical solutions, as shown in Fig. \ref{fig:erf-fourier2} in the Appendix.

\begin{figure}
    \centering
    \includegraphics[width=0.5\columnwidth,clip=true]{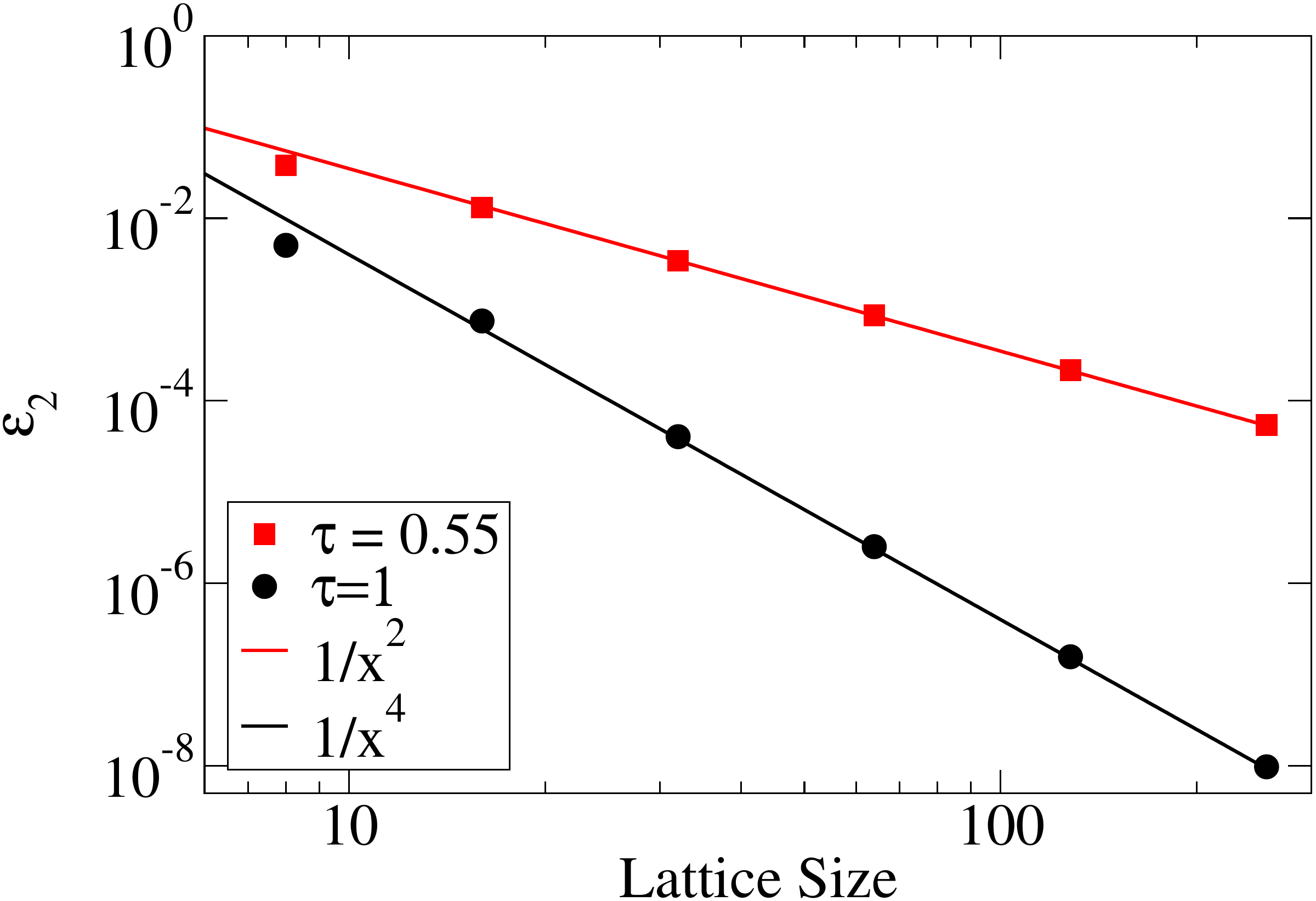}
    \caption{Convergence of our simulation results for different lattice sizes. For pairs $(\tau,\theta)$ for which we have $\alpha=0$, we obtain fourth-order convergence, whereas we obtain second-order convergence where $\alpha \neq 0$ as predicted by our theory.}
    \label{fig:convergence}
\end{figure}

There is another prediction we can obtain from our derivation of the correction term. In Figure \ref{fig:ratio-density}, we see bold black lines indicating the domain for $\alpha(\tau,\theta)=0$. For these values, our theory predicts that the original method (derived only to second order) is actually a fourth-order method. To test this prediction, we choose $\tau=1$ and $\theta=1/3$ for which we have $\alpha=0$, and examine the convergence of the method to the Fourier term analytical solution $\epsilon_2$ of Eqn. (\ref{eqn:eps}). In particular, we examine the same periodic system as before by choosing $F=0.0576$ for different lattice sizes. Keeping $F$ constant implies that the number of iterations scales as the square of the lattice size, which is sometimes known as diffusive scaling. This is shown in Figure \ref{fig:convergence}, where we see that we indeed find fourth-order convergence when $\alpha=0$. For pairs of $\tau$ and $\theta$ for which we have nonzero correction terms (\text{e.g.} we show the case $\tau=0.55$ and $\theta=1/3$), we see that we have a second-order convergence instead.

\section{Conclusions} \label{sec:conclusions}
In this paper we have examined whether a diffusive lattice Boltzmann method is an effective tool for examining problems related to Fickian water diffusion in barrier coatings. This validation was assisted by our ability to derive an analytical solution for a simple, but not trivial, coatings problem. In Sec. \ref{sec:coating} we presented a real-space solution for the water content of a dry coating that is initially exposed to a constant moisture reservoir on the surface. A second analytical solution in terms of Fourier components was presented in Sec. \ref{sec:fourier} that can be used both for the standard Fickian diffusion case already examined in Sec. \ref{sec:coating}, as well as the more complex fourth-order diffusion equation we derived as part of a higher-order hydrodynamic limit of the lattice Boltzmann equation. The two equivalent analytical solutions differ slightly because our Fourier series corresponds to a discrete system with only a finite number of Fourier terms.

For a simple initial implementation of the inlet boundary, we found excellent agreement only for a relaxation time $\tau=1$. Our analysis revealed that the disagreement for $\tau \neq 1$ was caused by assuming an equilibrium distribution as the reservoir boundary condition. Eventually we were able to define a ``perfect" boundary condition by doing away with the boundary altogether through an embedding of the system in a large periodic system that only requires periodic boundary conditions.

Along the way of our examination, we discovered that we can indeed identify a fourth-order accurate hydrodynamic limit of the diffusion equation. However, this higher-order correction was found to be irrelevant for the coatings problem considered here, as we could only identify a small region in parameter space where the fourth-order predictions were significantly more accurate. This may act as a cautionary tale that validating a higher-order correction does not guarantee that such predictions will always be more accurate for specific applications.

However, for the best cases, the numerical solutions agree with our analytical solutions almost as well as the two analytical solutions agree with each other, suggesting that the proposed method is indeed an excellent candidate to be applied to coatings problems.

In the future we expect to extend this from one-dimensional coatings problems, corresponding to full immersion of the coating, to the more complicated problem of droplets sitting on a coating. This case will require a full three-dimensional simulation, and wetting and drying problems then occur in one and the same simulation, spatially separated. Furthermore a droplet sitting on a coating would add a pressure gradient caused by the Laplace pressure in the drop. This may lead to an additional transport mode of advection, driven by the pressure gradient. This will require an extension of the current model to allow for some amount of advection as well. This advection would be expected to be highly overdamped, so that the local advection velocity would be simply proportional to the local pressure gradient. Technically doing this will require a replacement of the equilibrium distribution to one which allows for a non-zero first moment. 
Such simulations will significantly extend the current state of the art for coatings research which remains firmly focused on one-dimensional problems. 

\section*{Acknowledgments}
The authors thank Prof. Stuart Croll, Kent Ridl, and Reza Parsa of North Dakota State University for ongoing discussions and helpful insights. The second author was supported by the Strategic Environmental Research and Development Program under contract W912HQ-15-C-0012. Views, opinions, and/or findings contained in this report are those of the authors and should not be construed as an official Department of Defense position or decision unless so designated by other official documentation.

\appendix*
\section{Continuous and discrete solutions}
The error function solution in Eqn. (\ref{eqn:rho}) solves the second-order diffusion equation with the given boundary conditions in continuous real space (see \cite{crank1979mathematics} for a basic form of the derivation). However, we later compute a solution by transforming the appropriate initial condition into Fourier space, performing a second-order time evolution, and then transforming back into real space. This process uses a finite number of $k$ modes in each transform, and necessarily implies a discrete lattice sampling of both the initial condition in real space and the time-evolved form in Fourier space. We therefore expect a discrepancy when directly comparing the two solutions: the first is a solution to the continuous diffusion equation that is examined at discrete lattice points for comparison to the simulation, while the second is a sampled solution to the discrete lattice diffusion equation, the continuous form of which would require (in theory) an infinite number of $k$ modes to match the continuous case.

\begin{figure}
    \centering
    \includegraphics[width=0.5\columnwidth,clip=true]{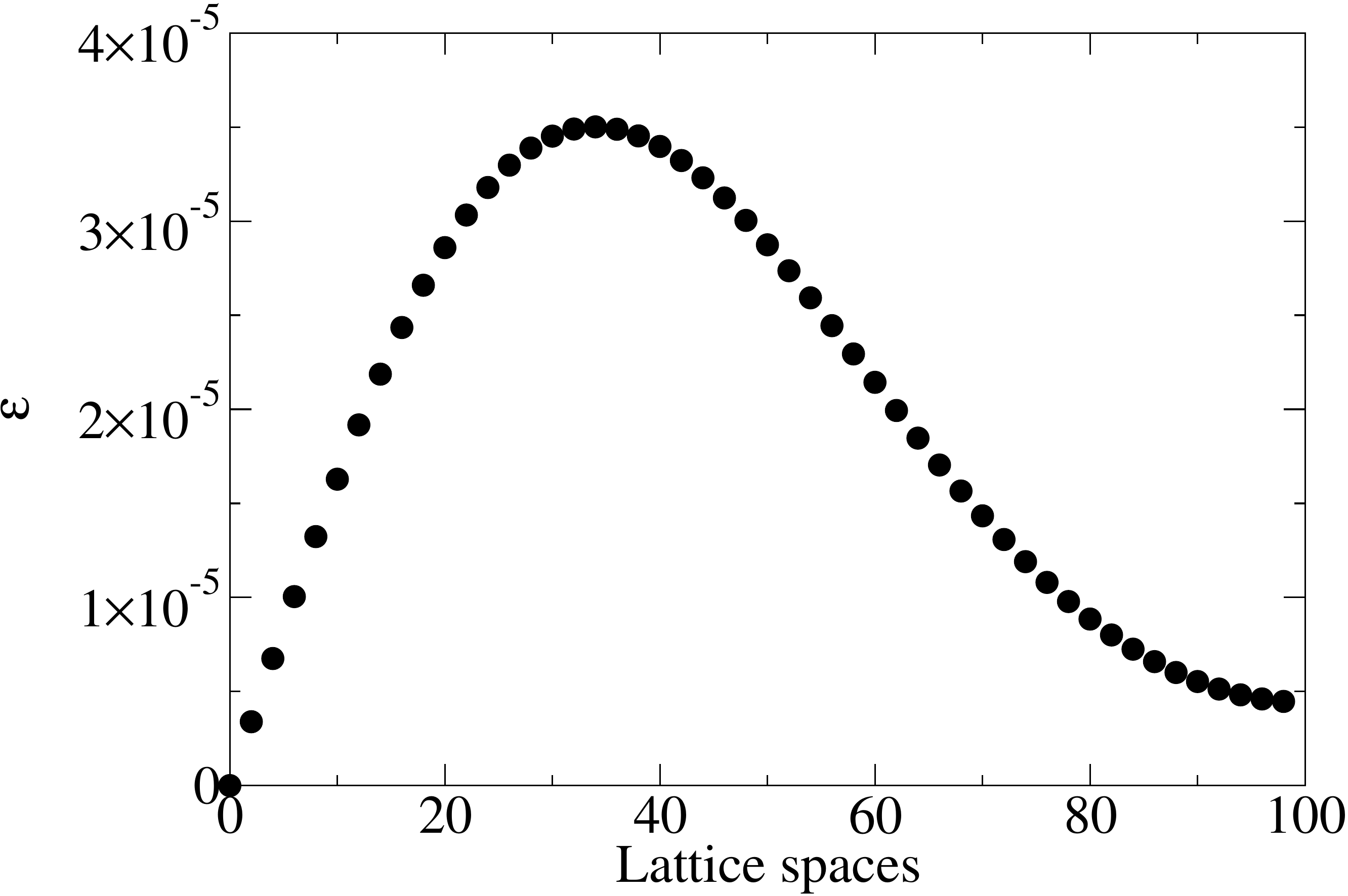}
    \caption{Absolute error profile $\epsilon$ between second-order error function and second-order Fourier solutions to the continuous and discrete diffusion equation, respectively.}
    \label{fig:erf-fourier2}
\end{figure}

To examine the extent to which these solution forms differ from each other, we compute both at the same scaled four-hour time at each lattice site, and plot the absolute value of the difference, $\epsilon$, in Fig. \ref{fig:erf-fourier2}. The two solutions agree to within $10^{-5}$ of each other. Since this error is on the order of the remaining error for the periodically-embedded simulation, we conclude that any further correction of simulation results renders any error obscured by differences between the discrete and continuous solutions to the diffusion equation, and is of no practical consequence.

\bibliographystyle{plain}
\bibliography{main}

\begin{thebibliography}{10}

\bibitem{Baukh20123304}
Viktor Baukh, Hendrik~P. Huinink, Olaf~C.G. Adan, Sebastiaan~J.F. Erich, and
  Leendert~G.J. van~der Ven.
\newblock Predicting water transport in multilayer coatings.
\newblock {\em Polymer}, 53(15):3304 -- 3312, 2012.

\bibitem{crank1979mathematics}
John Crank.
\newblock {\em The Mathematics of Diffusion}.
\newblock Oxford University Press, 1979.

\bibitem{de1998monitoring}
L~De~Rosa, T~Monetta, DB~Mitton, and F~Bellucci.
\newblock Monitoring degradation of single and multilayer organic coatings {I}.
  absorption and transport of water: Theoretical analysis and methods.
\newblock {\em Journal of the Electrochemical Society}, 145(11):3830--3838,
  1998.

\bibitem{PhysRevLett.56.1505}
U.~Frisch, B.~Hasslacher, and Y.~Pomeau.
\newblock Lattice-gas automata for the navier-stokes equation.
\newblock {\em Phys. Rev. Lett.}, 56:1505--1508, Apr 1986.

\bibitem{PREGinzburg}
Irina Ginzburg.
\newblock Prediction of the moments in advection-diffusion lattice boltzmann
  method. i. truncation dispersion, skewness, and kurtosis.
\newblock {\em Phys. Rev. E}, 95:013304, Jan 2017.

\bibitem{Kroll201582}
D.M. Kroll and S.G. Croll.
\newblock Influence of crosslinking functionality, temperature and conversion
  on heterogeneities in polymer networks.
\newblock {\em Polymer}, 79:82 -- 90, 2015.

\bibitem{LePRE2015}
Guigao Le, Othmane Oulaid, and Junfeng Zhang.
\newblock Publisher's note: Counter-extrapolation method for conjugate
  interfaces in computational heat and mass transfer [{P}hys. {R}ev. {E}
  \textbf{91} , 033306 (2015)].
\newblock {\em Phys. Rev. E}, 92:049904, Oct 2015.

\bibitem{LiPRE2014}
Like Li, Chen Chen, Renwei Mei, and James~F. Klausner.
\newblock Conjugate heat and mass transfer in the lattice {B}oltzmann equation
  method.
\newblock {\em Phys. Rev. E}, 89:043308, Apr 2014.

\bibitem{doi:10.1021/bk-2002-0805.ch001}
Jonathan~W. Martin.
\newblock {\em Repeatability and Reproducibility of Field Exposure Results},
  chapter~1, pages 2--22.
\newblock American Chemical Society, 2001.

\bibitem{AIC:AIC690480104}
Minas~M. Mezedur, Massoud Kaviany, and Wayne Moore.
\newblock Effect of pore structure, randomness and size on effective mass
  diffusivity.
\newblock {\em AIChE Journal}, 48(1):15--24, 2002.

\bibitem{Mu2007}
Deqiang Mu, Zhong-Sheng Liu, Cheng Huang, and Ned Djilali.
\newblock Prediction of the effective diffusion coefficient in random porous
  media using the finite element method.
\newblock {\em Journal of Porous Materials}, 14(1):49--54, 2007.

\bibitem{Pathania2017149}
Aman Pathania, Raj~Kumar Arya, and Sanjeev Ahuja.
\newblock Crosslinked polymeric coatings: Preparation, characterization, and
  diffusion studies.
\newblock {\em Progress in Organic Coatings}, 105:149 -- 162, 2017.

\bibitem{qian1992lattice}
YH~Qian, Dominique d'Humi{\`e}res, and Pierre Lallemand.
\newblock Lattice {BGK} models for {N}avier-{S}tokes equation.
\newblock {\em EPL (Europhysics Letters)}, 17(6):479, 1992.

\bibitem{SAHIMI19912225}
Muhammad Sahimi and Dietrich Stauffer.
\newblock Efficient simulation of flow and transport in porous media.
\newblock {\em Chemical Engineering Science}, 46(9):2225 -- 2233, 1991.

\bibitem{shan}
Xiaowen Shan and Gary Doolen.
\newblock Diffusion in a multicomponent lattice {B}oltzmann equation model.
\newblock {\em Phys. Rev. E}, 54:3614--3620, Oct 1996.

\bibitem{PhysRevLett.75.830}
Michael~R. Swift, W.~R. Osborn, and J.~M. Yeomans.
\newblock Lattice boltzmann simulation of nonideal fluids.
\newblock {\em Phys. Rev. Lett.}, 75:830--833, Jul 1995.

\bibitem{Taylor2012169}
S.R. Taylor, F.~Contu, R.~Santhanam, and P.~Suwanna.
\newblock The use of cationic fluoroprobes to characterize ionic pathways in
  organic coatings.
\newblock {\em Progress in Organic Coatings}, 73(2–3):169 -- 172, 2012.

\bibitem{PhysRevE.74.056703}
A.~J. Wagner.
\newblock Thermodynamic consistency of liquid-gas lattice {B}oltzmann
  simulations.
\newblock {\em Phys. Rev. E}, 74:056703, Nov 2006.

\bibitem{electrostatics}
A.J. Wagner and S.~May.
\newblock Electrostatic interactions across a charged lipid bilayer.
\newblock {\em Eur Biophys J}, 36:293--303, April 2007.

\bibitem{PhysRevE.94.033302}
Alexander~J. Wagner and Kyle Strand.
\newblock Fluctuating lattice {B}oltzmann method for the diffusion equation.
\newblock {\em Phys. Rev. E}, 94:033302, Sep 2016.

\bibitem{wolfdiffusion}
Dieter Wolf-Gladrow.
\newblock A lattice {B}oltzmann equation for diffusion.
\newblock {\em Journal of Statistical Physics}, 79(5-6):1023--1032, 1995.

\bibitem{Zee201555}
Malia Zee, Aaron~J. Feickert, D.M. Kroll, and S.G. Croll.
\newblock Cavitation in crosslinked polymers: Molecular dynamics simulations of
  network formation.
\newblock {\em Progress in Organic Coatings}, 83:55 -- 63, 2015.

\end{thebibliography}

\end{document}